\documentstyle{mn}

\newif\ifAMStwofonts

\input epsf

\def\plotone#1{\centering \leavevmode
\epsfxsize=\columnwidth \epsfbox{#1}}
\def\plottwo#1#2{\centering \leavevmode
\epsfxsize=.99\columnwidth \epsfbox{#1} \hfil
\epsfxsize=.99\columnwidth \epsfbox{#2}}
\def\plotthree#1#2#3{\centering \leavevmode 
\epsfxsize=.66\columnwidth \epsfbox{#1} 
\epsfxsize=.66\columnwidth \epsfbox{#2} 
\epsfxsize=.66\columnwidth \epsfbox{#3}}

\def\apj{ApJ}                 
\def\apjl{ApJ}

\def\aap{A\&A}

\def\mnras{MNRAS}

\def\pasj{PASJ}

\def\ssr{Space~Sci.~Rev.}     
                 
\def\nat{Nature}

\ifoldfss
  \ifCUPmtlplainloaded \else
    \NewTextAlphabet{textbfit} {cmbxti10} {}
    \NewTextAlphabet{textbfss} {cmssbx10} {}
    \NewMathAlphabet{mathbfit} {cmbxti10} {} 
    \NewMathAlphabet{mathbfss} {cmssbx10} {} 
  \fi
  \ifAMStwofonts
    \ifCUPmtlplainloaded \else
      \NewSymbolFont{upmath} {eurm10}
      \NewSymbolFont{AMSa} {msam10}
      \NewMathSymbol{\upi}     {0}{upmath}{19}
      \NewMathSymbol{\umu}     {0}{upmath}{16}
      \NewMathSymbol{\upartial}{0}{upmath}{40}
      \NewMathSymbol{\leqslant}{3}{AMSa}{36}
      \NewMathSymbol{\geqslant}{3}{AMSa}{3E}

       \let\le=\leqslant
       
    \fi
  \fi
\fi 

\ifnfssone
  \newmathalphabet{\mathit}
  \addtoversion{normal}{\mathit}{cmr}{m}{it}
  \addtoversion{bold}{\mathit}{cmr}{bx}{it}
  \newmathalphabet{\mathbfit} 
  \addtoversion{normal}{\mathbfit}{cmr}{bx}{it}
  \addtoversion{bold}{\mathbfit}{cmr}{bx}{it}
  \newmathalphabet{\mathbfss} 
  \addtoversion{normal}{\mathbfss}{cmss}{bx}{n}
  \addtoversion{bold}{\mathbfss}{cmss}{bx}{n}
  \ifAMStwofonts
    \ifCUPmtlplainloaded \else
      \UseAMStwoboldmath
      \makeatletter
      \new@mathgroup\upmath@group
      \define@mathgroup\mv@normal\upmath@group{eur}{m}{n}
      \define@mathgroup\mv@bold\upmath@group{eur}{b}{n}
      \edef\UPM{\hexnumber\upmath@group}
      \new@mathgroup\amsa@group
      \define@mathgroup\mv@normal\amsa@group{msa}{m}{n}
      \define@mathgroup\mv@bold\amsa@group{msa}{m}{n}
      \edef\AMSa{\hexnumber\amsa@group}
      \makeatother
      \mathchardef\upi="0\UPM19
      \mathchardef\umu="0\UPM16
      \mathchardef\upartial="0\UPM40
      \mathchardef\leqslant="3\AMSa36
      \mathchardef\geqslant="3\AMSa3E

       \let\le=\leqslant

    \fi
  \fi
\fi 

\ifnfsstwo
  \DeclareMathAlphabet{\mathbfit}{OT1}{cmr}{bx}{it}
  \SetMathAlphabet\mathbfit{bold}{OT1}{cmr}{bx}{it}
  \DeclareMathAlphabet{\mathbfss}{OT1}{cmss}{bx}{n}
  \SetMathAlphabet\mathbfss{bold}{OT1}{cmss}{bx}{n}
  \ifAMStwofonts
    \ifCUPmtlplainloaded \else
      \DeclareSymbolFont{UPM}{U}{eur}{m}{n}
      \SetSymbolFont{UPM}{bold}{U}{eur}{b}{n}
      \DeclareSymbolFont{AMSa}{U}{msa}{m}{n}
      \DeclareMathSymbol{\upi}{0}{UPM}{"19}
      \DeclareMathSymbol{\umu}{0}{UPM}{"16}
      \DeclareMathSymbol{\upartial}{0}{UPM}{"40}
      \DeclareMathSymbol{\leqslant}{3}{AMSa}{"36}
      \DeclareMathSymbol{\geqslant}{3}{AMSa}{"3E}

       \let\le=\leqslant

    \fi
  \fi
\fi 

\ifCUPmtlplainloaded \else
  \ifAMStwofonts \else 
    \def\upi{\pi}
    \def\umu{\mu}
    \def\upartial{\partial}
  \fi
\fi

\title[Equivalent width of the iron fluorescent line]{Equivalent width, shape and proper motion of the iron
fluorescent line emission from the 
molecular  clouds as an indicator of the illuminating source X-ray
flux history}

\author[Sunyaev \& Churazov]{R.~Sunyaev,$^{1,2}$ E.~Churazov,$^{1,2}$\\
$^1$ MPI fur Astrophysik, Karl-Schwarzschild-Strasse 1, 85740
Garching, Germany \\
$^2$ Space Research Institute (IKI), Profsouznaya 84/32, Moscow 117810, 
Russia}

\date{To appear in MNRAS}

\pagerange{\pageref{firstpage}--\pageref{lastpage}}
\pubyear{1998}

\begin{document}

\maketitle

\label{firstpage}
\begin{abstract}
Observations of the diffuse emission in the 8--22 keV energy range, elongated
parallel to the Galactic plane \cite{smp93} and 
detection of the strong 6.4 keV fluorescent line with $\sim$ 1 keV
equivalent width from some 
giant molecular clouds (e.g. Sgr B2) in the Galactic Centre region
\cite{koy94} suggest that the neutral matter of these clouds is (or
was) illuminated by powerful X-ray radiation, which gave rise to the
reprocessed 
radiation. The source of this radiation remains unknown.  Transient
source close to the
Sgr B2 cloud or short outburst of the X-ray emission
from supermassive black hole at the Galactic Centre are the two prime
candidates under consideration. We argue that new generation of X-ray
telescopes combining 
very high sensitivity  and excellent energy and angular resolutions
would be able to discriminate between these two possibilities studying
time dependent changes of the morphology of the surface brightness
distribution, the equivalent width and the shape of the
fluorescent line in the Sgr B2 and other molecular clouds in the region.
We note also that detection of broad and complex structures near the
6.4 keV line in
the spectra of distant AGNs, which are X--ray weak now, may
prove the presence of violent activity of the central engines of these objects
in the past. Accurate measurements of the line shape may provide an
information on the time elapsed since the outburst. Proper motion
(super or subluminal) of the fluorescent radiation wave front can give
additional information on the location of the source. Observations of
the described effects can provide unique information on the matter
distribution inside Sgr B2 and other giant molecular clouds.
\end{abstract}

\section{Introduction}
Prediction \cite{smp93} and discovery \cite{koy94,koy96} of the
bright iron fluorescent $K_\alpha$ line in the direction of the
molecular cloud Sgr B2 and 
Radio Arc in the Galactic Centre region should not remain unnoticed by
the astrophysicists planning in the nearest future launch of the
sensitive X--ray spectrometers  on board {\it AXAF, XMM,
ASTRO--E, ABRIXAS, Spectrum--X--Gamma}. These spectrometers will
provide high angular 
resolution from seconds to minutes of arc and spectral resolution from
5 to 140 eV near X-Ray lines of iron in the 6--7 keV energy band. The
missions of the next millennium, starting with 
{\it Constellation} (White, Tananbaum \& Kahn, 1997) and
{\it XEUS} \cite{tur97}, are to achieve energy resolution of 2 eV and
better. Particularly 
relevant problem, which deserves further consideration,  is the
illumination of a molecular hydrogen cloud with column density
$N_H\sim 10^{23} - 10^{24}~~cm^{-2}$ (i.e. $\tau_T\sim 
0.1-1$) by a variable X--ray emission from a bright transient
source inside the cloud or outside the cloud (e.g. short episode of
the effective accretion onto Sgr A* due to the tidal star
disruption). The solution of this problem 
allows one to study the time evolution of the spectrum emerging from
the cloud after fading of the primary X--ray source. The radius of the
Sgr B2 cloud is of the order of 20 pc (e.g. Lis \& Goldsmith 1989),
although the  
size of the dense core(s) is significantly smaller ($\sim 0.3$ pc,
e.g. de Vicente et al. 1997). Depending on the mutual 
location of the cloud and a primary source of the continuum emission
substantial evolution of the morphology, flux, equivalent width and
shape of the iron fluorescent line might be noticed on the time scale as
short as 0.1--10 years (the value which is not incomparable with the 
life time of the best modern space observatories). Three years already passed
since the moment of the first firm detection of the line from Sgr B2
\cite{koy94,koy96}. Detailed observations would allow one to reveal
the geometry of the problem (i.e. mutual location of the primary source and the
cloud), time elapsed since fading of the primary source
flux. Observations might also shed additional light on the mass of the cloud,
it's uniformity and, provided energy resolution better than 1 eV, on
the matter and velocity distribution inside the cloud. 

As the first approximation we are considering below qualitatively how
the effects of time delay, large opacity and scattering by bound
electrons affect the appearance of the 6.4 keV line from a
molecular cloud illuminated by a continuum X--ray emission. Throughout
the paper we adopted the approximation of Morrison and McCammon 
(1983) for photoelectric absorption ($\sigma_{ph}(E)$) in the neutral
gas, having a normal abundance of 
heavy elements, an abundance of iron of $\delta_{Fe}=3.3\times 10^{-5}$ with
respect to hydrogen, a cross section of 
photoabsorption from iron K-shell as $\sigma_{Fe}(E)=3.53\times
10^{-20}\times (E/7.1~keV)^{-3} cm^2/atom$ and a $K_\alpha$ fluorescent
yield $Y=0.3$ (e.g. Bambinek et al., 1972).  For simplicity we are considering
(unless stated otherwise) the 
simplest case  of a point source of continuum X-ray emission in the centre of a
spherically symmetric cloud of neutral gas.

The formation of the $K_\alpha$ line from the neutral matter
illuminated by a continuum radiation was considered in many
publications (e.g. Basko, Sunyaev \& Titarchuk 1974, Fabian 1977,
Basko 1978, Vainshtein \& Sunyaev 1980, Inoue 1985, George \& Fabian
1991, Matt et al. 1991, Awaki et al. 1991, Nandra \& George 1994,
Ghisellini et al. 1994). In the discussion below we present the
arguments which are particularly relevant to the observations of the
Galactic Centre region in the neutral iron fluorescent line.

This paper does not pretend to explain the nature of an X--ray
emission from the Galactic Centre region. Instead we consider a number
of simple effects which might play an important role in the environment like
the central region of our Galaxy. These effects could be used
by the missions like {\it Constellation} and {\it XEUS} to verify the hypothesis that
molecular clouds near the Galactic Centre were exposed to outburst of
hard X--Ray radiation:
\begin{itemize}
\item Dependence of the morphology of the 6.4 keV surface brightness
distribution on the mutual location of the source and the cloud.
\item Apparent motion (sub or superluminal) of the features associated
with propagation of the source radiation through the clouds.
\item Evolution of equivalent width and shape of the 6.4 keV line as
an indicator of multiple scatterings and time elapsed since outburst.
\end{itemize}

The structure of the paper is as follows: in sections 2 and 3 we
discuss temporal behavior of scattered flux and morphology of the
scattered radiation associated with first scattering, in the
subsequent sections 4--7 we argue that 
equivalent width and the shape of the fluorescent line (due to
multiple scatterings) may be used as another indicator of  cloud
illumination with powerful flares in the past, section \ref{ssum}
summarizes the results. In appendix a simplified derivation of the
evolution of the equivalent width and shape of the line after multiple
scatterings is given.

\section{The luminosity of illuminating source}
\label{smor}
      Observations of the Galactic Centre region with the GRANAT/ART--P
telescope revealed a diffuse component above $\sim$ 8 keV elongated
parallel to the Galactic plane and resembling the distribution of the
molecular gas clouds \cite{smp93,msp93}. It was suggested that this
component is due to the Thomson scattering by dense molecular gas of the
X-rays from nearby compact sources and as a consequence existence of
bright $K_\alpha$ iron fluorescent line was proposed. Further evidence
in support of this 
assumption came from ASCA observations of this region, which revealed
strong 6.4 keV line flux from the general direction of the largest molecular
complexes (e.g. Koyama, 1994). The Sgr B2 complex  was found to be
especially bright in the 6.4 keV line, at the level of $1.7~10^{-4}~
photon~s^{-1} cm^{-2}$ \cite{koy96}.  Such a high flux in the fluorescent line
requires a powerful source of primary X-ray emission which gives rise
to the reprocessed radiation. Since the light 
crossing time of the Galactic Centre region is as long as several hundred
years, observed reprocessed radiation may  be associated with the source which
is dim at present, but was very bright some hundred years ago. Such a
scenario has been suggested by Sunyaev, Markevitch \& Pavlinsky
\shortcite{smp93} and Koyama \shortcite{koy94}, Koyama et
al. \shortcite{koy96}, 
the Galactic Centre (i.e. Sgr A$^*$) itself being the primary
candidate. Given the distance from the Sgr B2 complex 
to the GC ($\sim 40'$ or about 100 pc in projection) Sunyaev,
Markevitch \& Pavlinsky \shortcite{smp93} and Koyama et
al. \shortcite{koy96} estimated the 
luminosity of the putative GC source in excess of $10^{39}~erg~s^{-1}$ some
hundred years ago. The flux in the 6.4 keV line from the cloud exposed
to the continuum radiation is given by the expression:
\begin{eqnarray}        \label{e1}
F_{6.4}= \frac{\Omega}{4\pi D^2}~ n_{Fe} r Y
~\int_{7.1}^{\infty}{I(E)\sigma_{ph}(E)dE}~~~phot~s^{-1} cm^{-2}
\end{eqnarray}
where $\Omega$ is the solid angle of the cloud from the location of
the primary source, $D$ is the distance to the observer, $n_{Fe} r$ is
iron column density of the cloud, $I(E)$ is the
photon spectrum of the primary source ($phot~s^{-1} keV^{-1}$). Since
$\sigma_{ph}(E)$ is a steep function of energy the 6.4 keV flux
depends mainly on the source flux at $\sim$ 7--9 keV.  It is
convenient therefore to 
express 6.4 keV flux through the source flux at 8 keV:
\begin{eqnarray}        \label{e2}
F_{6.4}= \phi~ \frac{\Omega}{4\pi D^2}~\frac{\delta_{Fe}}{3.3\times
10^{-5}}~~\tau_T ~I(8~keV)~~~phot~s^{-1} cm^{-2}
\end{eqnarray}
where $\phi$ is a factor of the order of unity, which depends (weakly)
on the shape of the source spectrum. For the bremsstrahlung spectrum
factor $\phi$ varies from 1 to 1.3 when temperature is varying from 5
to 150 keV. For convenience one can replace $I(8)$ in the above expression by
the measure of the continuum source luminosity at 8 keV in the 8 keV wide
energy band: $L_8=I(8)\times 8\times 8 \times 1.6~10^{-9}~erg~s^{-1}$.
\begin{eqnarray}        \label{e3}
F_{6.4}= \phi~ 10^{7} \frac{\Omega}{4\pi
D^2}~\frac{\delta_{Fe}}{3.3\times 10^{-5}}~~\tau_T L_8~~~phot~s^{-1} cm^{-2}
\end{eqnarray}
The value of $L_8$ characterizes the luminosity of the source in
standard X--ray band. E.g. for bremsstrahlung spectra with
temperatures from 5 up to 150 keV $L_8$ corresponds to $\approx$
40--45 per cent of the source luminosity in the 1--20 keV band. 
Thus the luminosity of the source required to produce
observed 6.4 keV flux is:
\begin{eqnarray}        \label{lumout}
L_8\approx 6~10^{38}\times \left ( \frac{F_{6.4}}{10^{-4}} \right ) \times
\left ( \frac{0.1}{\tau_T} \right ) \times \left (
\frac{\delta_{Fe}}{3.3\times 10^{-5}} \right )^{-1} \nonumber \\
\times \left (\frac{R}{100~pc}\right )^2
\end{eqnarray} 
where $R$ is the distance from the source to the cloud. The above crude
estimate assumes that the source is well outside the cloud, which has
diameter of $\sim$14 pc \cite{koy96} and $\tau_T\ll 1$. Although
high enough this 
value is still much below the Eddington limit of $\sim 10^{44}~erg~s^{-1}$ for
the $\sim 10^6 M_\odot$ black hole \cite{gen94} and even rather
short (e.g. several days) flare at  
the Eddington level could provide the required flux. Note that if the
duration of the flare is shorter that the light crossing time of
the cloud the above estimate should be multiplied by a factor roughly
$\frac{r/c}{\Delta t}$, where $r/c$ is the light crossing time of the
cloud and $\Delta t$ is the duration of the flare. In other words for
very short flare the product $L\times\Delta t$ (i.e. luminosity
$\times$ duration) defines the 6.4 keV flux \cite{smp93}.
 A less energetic object is
required if one assumes that the primary source of continuum emission is
located in the vicinity of the Sgr B2 complex (or even inside it). For
the source embedded into the uniform cloud the luminosity is only: 
\begin{eqnarray}        \label{lumin}
L_8 \sim 6~10^{35}\times \left ( \frac{F_{6.4}}{10^{-4}} \right ) \times
\left ( \frac{0.1}{\tau_T} \right ) \times \left (
\frac{\delta_{Fe}}{3.3\times 10^{-5}} \right )^{-1}
\end{eqnarray}
For hard spectra (e.g. bremsstrahlung with $kT\sim100$ keV) the 1--150
keV luminosity is factor of $\sim 7$ larger than $L_8$, but it
is still well consistent with observed luminosities of X--ray Novae with hard
spectra. This estimate should also be increased if the source was
bright during period of time shorter than the light crossing time of the cloud.

In the presence of the steady
primary source inside the cloud the expected equivalent width of the 6.4
keV line (with respect to the primary continuum flux) $EW$ is about
$1\times \tau_T$ keV, where $\tau_T$ is the Thomson depth 
of the medium, surrounding the source. If the primary source radiation
diminishes (or if the source is outside the region over which the spectrum is
collected), then the equivalent width is equal to $\sim 1 $
keV (e.g. Fabian 1977, Vainshtein \& Sunyaev 1980)  -- the value,
which is rather insensitive 
to the cloud parameters like Thomson optical depth (as long as it is
low enough) or particular distribution of scattering matter. This
happens because both 6.4 
keV and scattered continuum fluxes are proportional to the optical
depth of the cloud and intensity of the primary radiation. In
particular case of the Sgr B2 cloud the equivalent width of the
observed line is of the
order of 1 keV \cite{koy96}, suggesting that observed spectrum is
dominated by scattered radiation. Provided that the cloud is optically
thin and abundance of iron is $\sim 3.3\times 10^{-5}$ then large value of
the equivalent width immediately excludes the possibility that steady
source of continuum radiation is located inside the cloud or is projected onto
the cloud. The hypothesis that hot plasma surrounding the cloud is
responsible for primary radiation can be rejected for the same reason
even without knowledge of it's luminosity. 
Furthermore assuming again that the source is steady and has hard
($T_e=150$ keV) bremsstrahlung spectrum then one would expect the
source to be seen in the GRANAT/SIGMA observations. Indeed for given
source spectrum one can estimate 35--100 keV luminosity 
of the source from equation (\ref{lumin}):
$L_{hard}\sim 3~10^{36}\times \left ( \frac{0.1}{\tau_T} \right )
\times \left ( \frac{\delta_{Fe}}{3.3\times 10^{-5}} \right
)^{-1}~erg~s^{-1}$. Actual conservative upper limit is at least factor of 5
lower (see Gilfanov \& Sunyaev, 1998). 

Another possibility is that X-ray emission of the Sgr B2 cloud is
produced by a number of steady sources (may be diffuse) spreaded across the
cloud. At  each particular energy only regions which are screened from
us by an absorbing depth of $\tau_{ph}(E) \le 1$ will contribute to the
emergent flux. As long as line of sight depths to different emitting regions
spans wide range of values, `flat' continuum spectrum, like observed
from Sgr B2 \cite{koy96}, can be formed. Observationally the spectrum
of such a cloud  might resemble the reflected spectrum, usually
considered in application to AGNs or X--ray binaries. However, in
order to provide large equivalent width of the 6.4 keV line, most of
the emitting regions should be screened from us by a hydrogen column of
few $10^{24}~cm^{-2}$. In this case (as discussed further in section
4) continuum photons at $\sim6$ keV will be absorbed, while photons
with higher energies (e.g. 10--20 keV) may reach periphery of the
cloud and produce 6.4 keV photons. The total luminosity of such a
cloud (after correction for absorption) should be considerably larger
than that given by expression (\ref{lumin}).

The constraints on the source luminosity can be drastically reduced if
we assume that the emission of the primary source is intrinsically strongly
collimated. This is particularly important when considering Sgr A* as
a potential source of powering the 6.4 keV line emission from Sgr B2. The
solid angle of the Sgr B2 cloud as from the Sgr A* position is of the
order or less than $\sim 0.02$ str. Therefore required flux from Sgr
A* (within the narrow cone) is only of the order of $L_8\sim
10^{36}~erg~s^{-1}$ i.e. almost 3 orders of magnitude lower than that given by
expression (\ref{lumout}). We note that this value places an absolute
lower limit 
on the luminosity of Sgr A* as a potential source for powering Sgr B2
cloud. The brightness of the Sgr B2 cloud would then be
explained by the location of the cloud within this narrow cone of the
Sgr A* emission. If this hypothesis is correct, than high angular resolution
observations could perhaps find evidence of the `jet--like'
structures formed by the Thomson scattering  (as discussed by
Gilfanov, Sunyaev \& Churazov 1987) by the molecular gas within the
cone of the emission. 

Another hypothesis of a similar kind is that we deal with obscured
AGN, i.e. that Sgr A* is a very bright steady X--ray source (with luminosity
as it follows from expression (\ref{lumout})), but cold matter (at least $N_H
\sim 10^{25}~cm^{-2}$) in the very vicinity of the source blocks our
line of sight (see e.g. Predehl \& Tr\"{u}mper 1994), while Sgr B2 is
exposed to the Sgr A* 
emission \cite{koy96}. The commonly accepted value for interstellar
extinction towards the brightest IR stars near the Galactic Centre is
$A_v\sim 31$ \cite{rrp89}, which corresponds to hydrogen column density of
$N_H\sim 6~10^{22}~cm^{-2}$ i.e. at least two orders of magnitude
lower, than required. Absorption by the partly ionized gas in the
shell surrounding Sgr A* is also low (Beckert et al. 1996). It is
possible, however, that the edge of the inner accretion disk obscures
our line of sight. The major problem with this hypothesis (see
e.g. Koyama et al., 1996) is the lack of bright infrared source at the
location of Sgr A* \cite{mre97}. 

Thus there is no obvious candidate for the source of primary radiation
which present day brightness is sufficient to power 6.4 keV emission from
Sgr B2 cloud. We therefore consider below the situation when the
source of continuum radiation is variable and it is now in the
weak state. One can consider two different possibilities for the
location of such a source: 
\begin{itemize}
\item[{\bf A}] The primary (transient) source was located inside (or very
close) to the Sgr B2 cloud. 
\item[{\bf B }] The primary radiation is due to intense emission from
Sgr A* in the past. 
\end{itemize}

In any case the emission of the source is variable and perhaps on the
time scales shorter than light crossing time of the cloud. E.g. for
typical X--ray Nova flux is falling down with e--folding time of the
order of 40 days (e.g. Tanaka \& Shibazaki, 1996), while the rising
time can be as 
short as one or few days. For the transient accretion
event onto Srg A* (e.g. capture of tidally disrupted star) we also
might expect a flare duration at a time
scale of years \cite{rees88}. Thus the observed picture may not be static
and some information can be obtained studying it's time evolution.

\section{Time dependent flux, morphology (bright spots) and proper
motion of the 6.4 keV line emitting regions}

\subsection{Single scattering problem}
Let us consider first the situation when the cloud is illuminated with the
relatively short flare. The geometry of the problem is schematically
shown in Fig.\ref{geom}. For the short flare surface of the parabola
denotes the positions with similar propagation times from the source to
the cloud and then to the (distant) observer. The surface
($z/c=(t^2-(x/c)^2)/2t$) has of course 
rotational symmetry over the $Z$ axis. Similar time/geometry
problem appears in many situations, e.g. while considering scattering of
the emission from central source by the electrons in hot intergalactic
media \cite{gsc87}, light echo from supernovae
(e.g. Chevalier, 1986) or even as an alternative to the 
gravitational lensing arcs \cite{katz87,mil87}. In connection with
Sgr B2 which one can make a number of predictions based on such a
picture.

\subsubsection{Bright spots}
If the flare is very short (much shorter than the light crossing
time of the cloud) then observed surface brightness at a given moment
will be determined not by the total optical depth of the cloud, but
rather by the density of the cloud at the surface of the parabola. The
parabola `scans' the density profile of the cloud. The surface
brightness is defined by the integral
$\int{\frac{I}{4\pi r^2}n dl}$ over the line of sight. The integration
limits are defined by two parabolas corresponding
to the beginning and the end of the flare. On can write a 
simple expression for the surface brightness (flux form the solid
angle $d\Omega$) of the 6.4 keV line
emission:
\begin{eqnarray}        \label{surf}
S=7~10^{-6} \left ( \frac{n}{10^5~cm^{-3}} \right ) \left (
\frac{\Delta t}{1~year}\right ) \left ( \frac{L_8}{10^{39}}\right )
\left ( \frac{100~pc}{x} \right ) ^2 \times \nonumber \\
\times \left ( \frac{d\Omega}{(15'')^2} \right )  
\frac{\eta^2}{(1+\eta^2)}~~~phot~s^{-1} cm^{-2}
\end{eqnarray}
where $\Delta t$ is the duration of flare, $\eta=x/ct$, $x$ is the
projected distance from the source to the bright spot, $t$ is
the time elapsed since the flare. The above formula (scaled to the
angular resolutions of the {\it XMM} and JET--X on {\it
Spectrum--X-Gamma}) shows that with integration 
time of $10^5$ seconds and with effective area from $\sim$ 300 to 3000
$cm^2$ at 
6.4 keV these instruments will be able to trace the density variations
in the cloud. The estimated size of dense condensations in
the Sgr B2 cloud of $\sim$0.5--0.3 pc (e.g. de Vicente et al., 1997)
is well matched 
with the angular resolution of these telescopes. Note that for search of
the bright spots energy resolution of typical CCD is sufficient.
Thus if the Sgr B2
cloud was indeed illuminated by the short flare, then one can expect
very strong variations (up to three orders of magnitude according to
the data on molecular lines tracers of high density) of the surface
brightness of the 6.4 keV flux 
across the cloud image on the angular scales, corresponding to the
size of the nonuniformities in the cloud (i.e. 10--20$''$). If on the
contrary the flare was sufficiently extended in time, then surface brightness
distribution will reflect total optical depth of the cloud (at a given
line of sight), which should be much smoother since contribution to
the total scattering mass from extended envelopes dominates. 

\subsubsection{Superluminal proper motion of the front}
\begin{figure}
\plotone{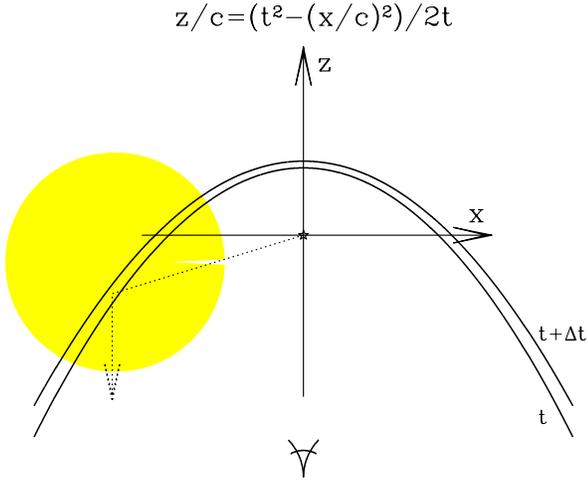}
\caption{
Each point at the surface of the parabola $z/c=\frac{t^2-(x/c)^2}{2t}$
has similar delay $t$ in propagation time from the source (marked with
asterisk) to the scattering place and then to the observer. In the
case of a short flare the fluorescent photons which are observed at
a given moment of time are 
produced in the neutral matter located at the surface of the
parabola. For a steady source which faded at a given moment of
time the exterior of the parabola is filled with primary continuum photons.
} 
\label{geom}
\end{figure}

Note that with time the 
position of the parabola will change and the velocity of this change
strongly depends on the mutual location of the source and the
cloud. In particularly `expansion' velocity of the parabola
(Fig.\ref{geom}) over X 
coordinate (at fixed Z) is $|\dot{x}/c|=1+\frac{(ct-|x|)^2}{2ct|x|}$,
i.e. $1\le | \dot{x}/c | < \infty$. Therefore it
is equal or larger than speed of 
light for any position. Similarly the velocity of parabola expansion
over Z coordinate (at fixed X) is
$\dot{z}/c=1-\frac{z}{ct}=0.5+\frac{x^2}{2(ct)^2}$ 
($1/2\le \dot{z}/c < \infty$). Note that $\dot{z} < c$ if $z > 0$ (the
cloud is behind the source) and $\dot{z} > c$ if $z < 0$ (the cloud is
in front of the source). Thus if the cloud is located closer to the
observer than the source  of the flare, then the whole cloud will be
`scanned' much quicker than the light crossing time of the
cloud. For the cloud behind the source time evolution will be
slowest. Stark et al. \shortcite{sta91} argue that since the radial
velocity of the Sgr 
B2 cloud relative to the GC is only $\sim$ 40 $km~s^{-1}$ the assumption that
it is on the circular orbit at $\sim$ 180 $km~s^{-1}$ would place it
$\approx$ 400 pc closer (or further away) from the 
observer than the Galactic Centre. It
is clear that in the case of short flare time dependent effects will
be very different for these two cases: for the nearer location one can
expect higher flux and much more rapid evolution. If on the contrary
the flare was very long (such that the whole cloud is filled with
photons), then these two cases will be virtually indistinguishable from each
other. However, taking into account that photoabsorption may
significantly affect the morphology (especially for dense cores,
having considerable depth) it might be possible two discriminate
between these two cases. 
\begin{figure}
\plotone{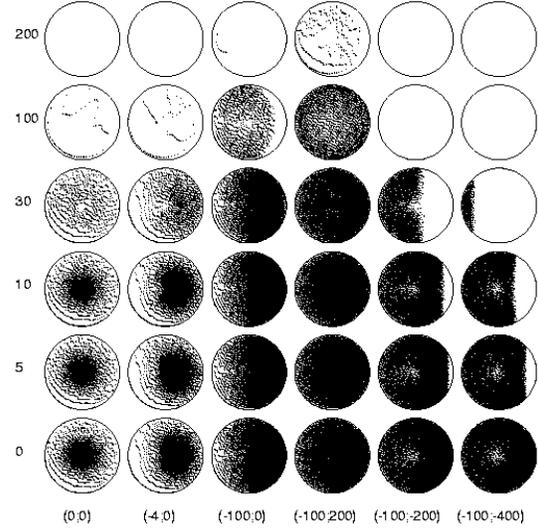}
\caption{
Morphology of the 6.4 keV line surface brightness distribution as a
function of time and relative position of the scattering cloud and the
compact source of continuum radiation. Time evolution assumes steady
source of continuum emission which faded at some moment of
time. The time tags marked in 
the left column indicate the time (in years) elapsed since the moment,
when the surface of the parabola (which exterior is filled by
primary radiation) 
touched the cloud for the first time (see text and Fig.1).  The position of the
cloud with respect to the compact source is indicated with a pair of
numbers (in parsecs) at the bottom of the figure. The radius of the
cloud was assumed to be 22.5 pc and density distribution according to
expression (\ref{nesgr}).} 
\label{mort}
\end{figure}

\subsubsection{Surface brightness distribution (steady source and long
flare)}
Shown in the Fig.\ref{mort} (bottom row of images) is the expected
surface brightness distribution of the spherically symmetric cloud
exposed to steady continuum radiation calculated for different positions of
the primary source. We assume here that density of hydrogen molecules
in the cloud has the following radial distribution: 
\begin{eqnarray}        \label{nesgr}
n(r)=2.2~10^3+\frac{7.7~10^4}{1+(\frac{r}{1.25~pc})^2}~cm^{-3},~~~~~~~~r<22.5~
pc
\end{eqnarray} 
This distribution closely resembles the model distribution derived by
Lis \& Goldsmith \shortcite{lg89}, 
although actual density distribution in the Sgr B2 complex is of course
much more complex (e.g. Hasegawa et al. 1994, de Vicente et
al. 1997). The surface brightness was estimated 
integrating over the line of sight the product of the gas density in the
cloud and radiation field density (which for steady source is
$\frac{I_0}{4\pi r^2}$, where $I_0$ is
the intensity of the compact source and $r$ is the distance from the
source). Photoabsorption and Thomson scattering were taken into
account. The position of the centre of the cloud with respect to the
primary source 
(in the plane defined by the cloud, source and observed as shown in
Fig.\ref{geom}) is 
indicated by the pair of numbers at the bottom of the
Fig.\ref{mort}. E.g. $(0;0)$ corresponds to the source at the centre of the
cloud; $(-100;200)$ corresponds to the cloud which is shifted by 100
pc to the left from the source and is located 200 pc further away from the
observer than the source. Other rows in Fig.\ref{mort} show time
evolution of the surface brightness morphology, assuming steady
source of continuum emission which faded at some moment of
time. The time tags marked in the left column indicate the time
elapsed since the moment, when the surface of the parabola (which
exterior is filled by primary radiation) touched the cloud for the
first time. The situation ({\bf A}) (source inside the cloud) is characterized
by a distinct peak in the surface brightness distribution. This peak
is simply due to the higher density of the primary radiation close to
the source ($I\propto \frac{I_0}{4 \pi r^2}$). For the source, located
at a 
large distance from the cloud (compared to the radius of the cloud)
surface brightness is much more uniform and reflects roughly Thomson
depth of the cloud on the line of sight. Photoabsorption and Thomson
scattering cause additional asymmetry -- the side of the cloud exposed
to the radiation should be brighter. As pointed out by Koyama et
al. \shortcite{koy96} 
the side of the Sgr B2 cloud towards Sgr A* is somewhat brighter,
perhaps suggesting that the primary source is at least in general direction of
Sgr A*. 

\begin{figure*}
\plottwo{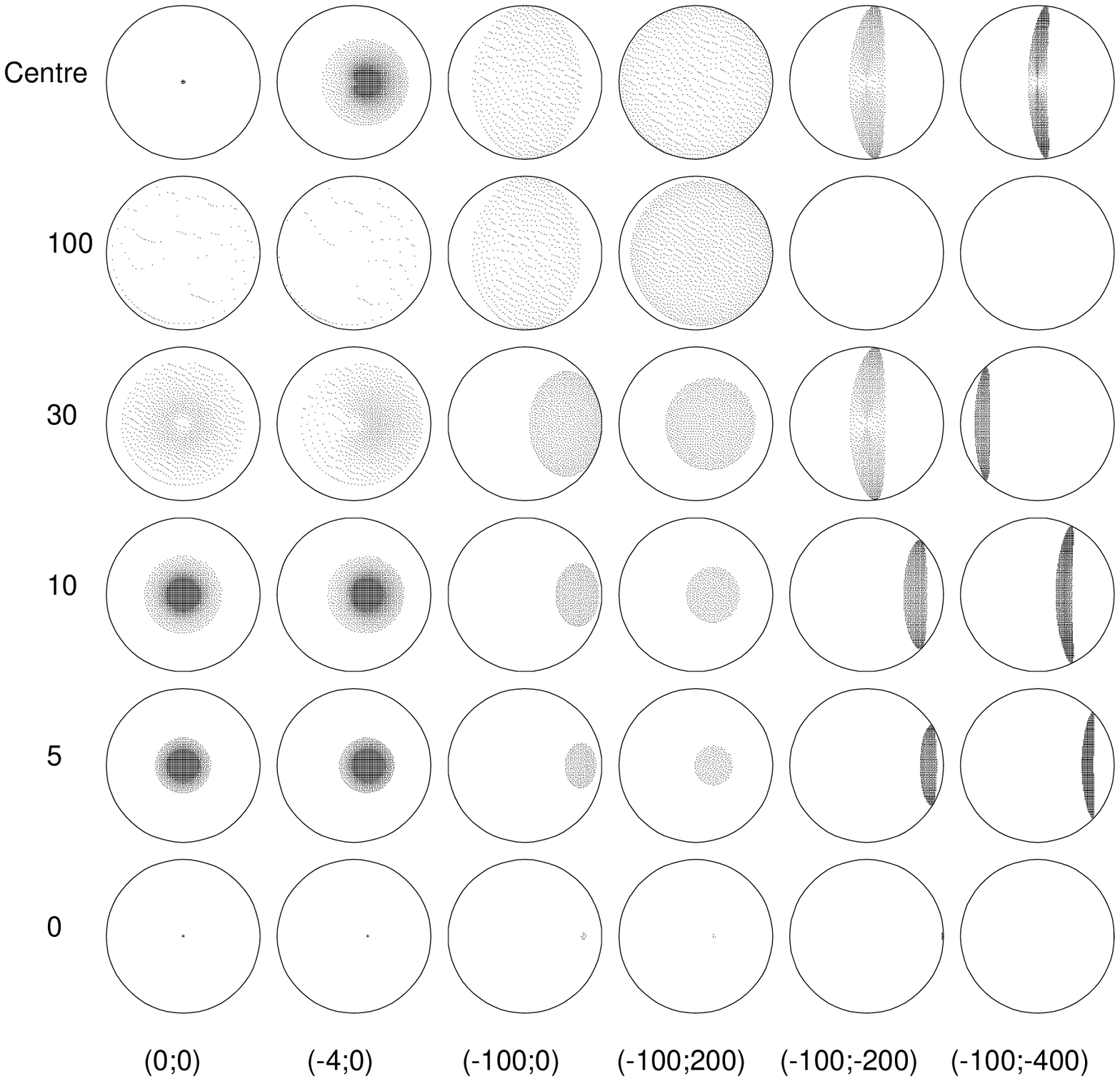}{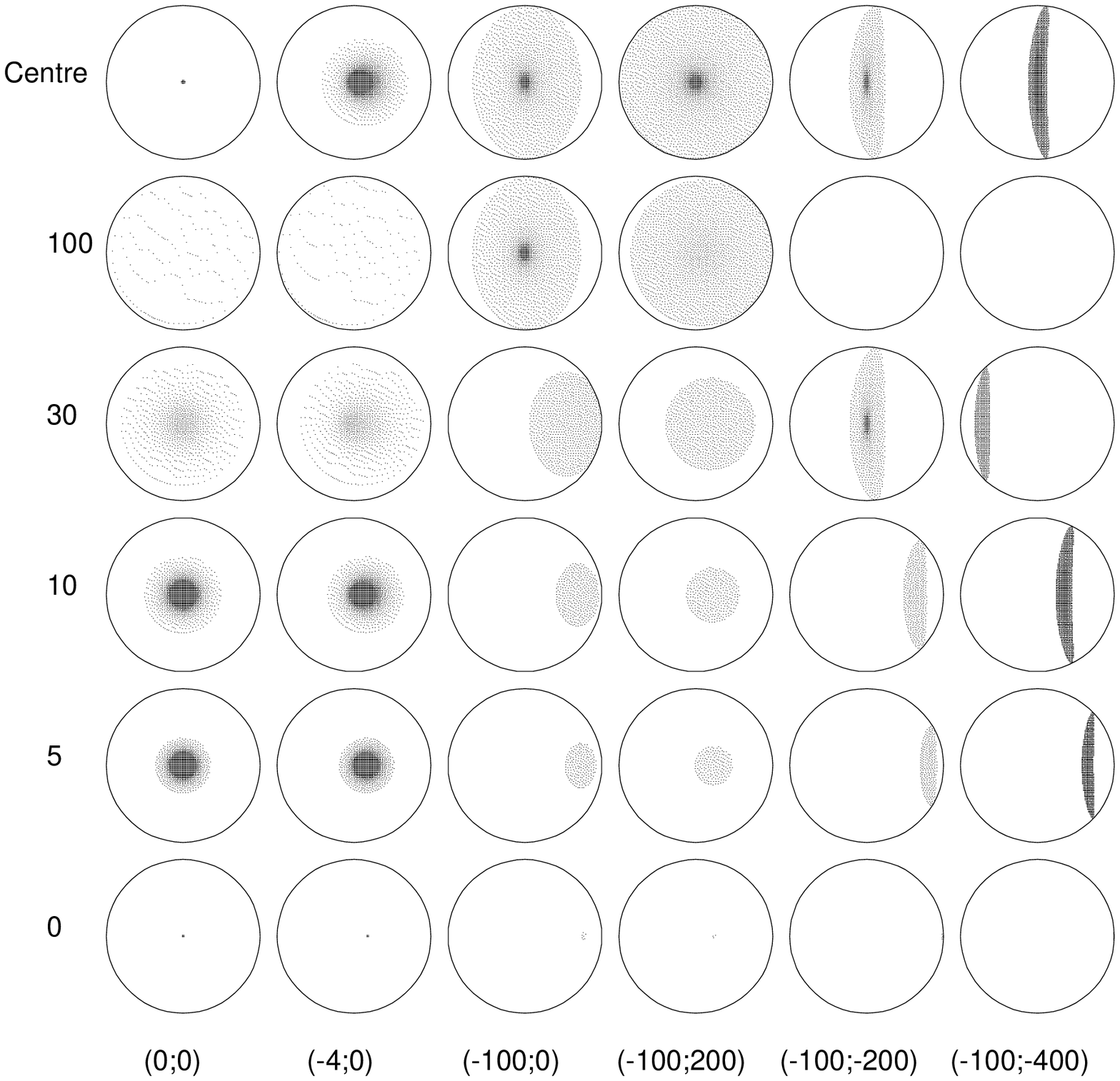}
\caption{
{\bf Left:} Morphology of the 6.4 keV line surface brightness distribution as a
function of time and relative position of the scattering cloud and the
compact source of continuum radiation. Time evolution assumes short
flare of continuum emission. The time tags marked in 
the left column indicate the time elapsed since the moment,
when the surface of the illuminating parabola touched the cloud for the first
time (see text and Fig.2). The
intensity of the the primary source was adjusted for each column in
order to have the same total flux from the cloud for row marked as
`30'(years). The position of the
cloud with respect to the source is indicated with a pair of
numbers at the bottom of the figure. The topmost row of images
corresponds to the moment of time when the surface of the parabola
goes through the centre of the cloud. Density distribution in the
cloud adopted from Lis and Goldsmith, 1989 (see equation
(\ref{nesgr})). {\bf Right:}  
The same as in the left figure, but for the 100 times less dense cloud
-- i.e. in the limit of optically thin cloud. Note that in this limit
surface brightness at any given moment of time reflects the density
distribution over the surface of the parabola.} 
\label{morf1}
\label{morf2}
\end{figure*}

\subsubsection{Surface brightness distribution (short flare)}
\begin{figure}
\plotone{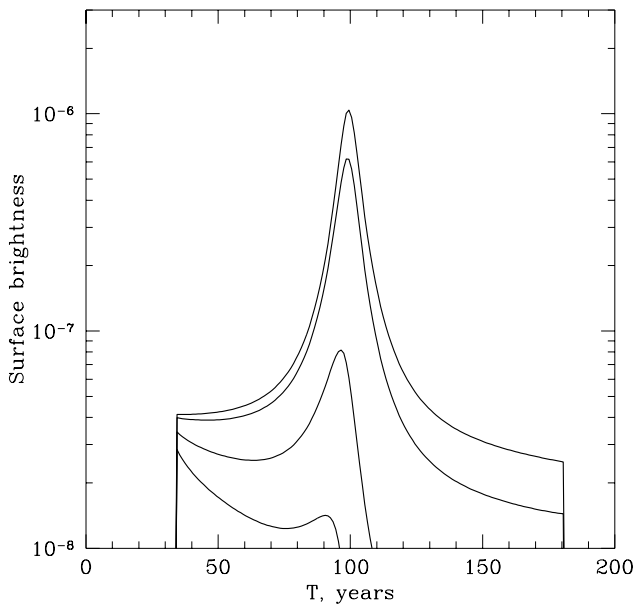}
\caption{
Time evolution of the surface brightness of the cloud towards the
centre. The shape of the density distribution is the same for all
curves, but normalizations are different. The lowest curve corresponds
to the Lis and Goldsmith (1989) model of the Sgr B2 cloud. The other
curves (from bottom to top) correspond to the cloud with density
scaled by factors of 0.5, 0.1 and 0.01 with respect to the Lis and
Goldsmith model. The flux of the source was scaled inversely to this
factor in order to keep the same intensity of the scattered flux
assuming optically thin regime. The primary 
source was assumed to be located at the same distance as the cloud,
but 100 pc away from the cloud in the picture plane.
} 
\label{lcsb}
\end{figure}

The Fig.\ref{morf1} shows similar time evolution of the surface
brightness for the case of a very short flare of the primary
source. Note that for an optically thin case (Fig.\ref{morf2}) surface
brightness distribution reflects the density distribution in the cloud
along the surface of the parabola. For the thicker cloud
(Fig.\ref{morf1}) photoabsorption may eliminate peaks associated with
denser condensations and even produce `holes' at their positions and
cast `shadows' on the more distant (from the primary source) parts
of the cloud. The temporal behavior of the surface brightness
distribution towards the centre of the cloud is shown in
Fig.\ref{lcsb}. Different curves corresponds to the different
normalizations of the density distribution in the cloud. For optically
thin cloud time evolution simply characterizes the density structure
of the cloud along the line of sight. For the thicker cloud
photoabsorption naturally limits the possibility to probe inner regions
of the cloud. 

\subsubsection{Time evolution of the flux from the whole cloud (short flare)}
\begin{figure}
\plotone{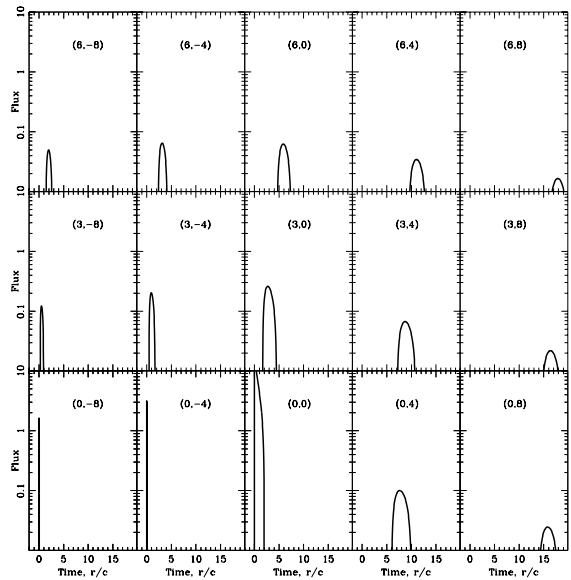}
\caption{Total flux from the cloud versus time for different mutual
locations of the source and the cloud. The duration of the flare was
assumed to be much shorter than the light crossing time of the
cloud. The positions indicated are measured in cloud radii. The time
is measured in units of light crossing time of the cloud.
} 
\label{flux_flare}
\end{figure}

.
The evolution of the total flux from the cloud (for the case of very
short flare) is shown in Fig.\ref{flux_flare}. The position of the
cloud with respect to the source $(x,z)$ is marked in each plot in units of
cloud radius $r$. Time is measured in units $r/c$. Bottom row
correspond to the case of the source, which projected position
coincides with the cloud centre. Upper two rows correspond to the
cloud which is shifted (in projection) by 3 and 6 cloud radii from the
source position. The relative distance of the source and the cloud
vary for the plots in different columns. The leftmost column correspond to the
cloud which is closer to the observer than the source by 8 cloud
radii. The rightmost column correspond to the cloud which if further
away than the source by 8 cloud radii.
One can see that duration of scattered flare strongly depends on the
position of the cloud with respect to the primary source. This
compensates to some extent the drop of the maximal flux with distance
(due to decrease of the solid angle) for the clouds located closer to
the observer than the source. The maximal flux drops with distance
along Z coordinate as $1/z$ for such clouds, compared to  $1/z^2$
dependence for clouds more distant, than the source.

Note that there are three factors limiting the time scale and the
amplitude of 6.4 keV flux variability from a given direction. First of
all the shortest possible time scale $\Delta t$ (and amplitude) are defined by
the variability of the primary source. The second factor is
nonuniformity of the matter distribution on the spatial scales
corresponding to the `thickness' of the parabola (which is
$0.5\Delta t c (1+\eta^2)$). The third factor is the velocity of the
parabola motion with respect to the spatial nonuniformities of the
neutral matter.

Finally one can note that although the flux of the 6.4 keV line may
vary by orders of magnitude depending on the geometry of the cloud and
it's relative position with respect to the source the equivalent width
should be approximately constant, since continuum is scattered in a
similar way. The equivalent width may vary however by a factor of
$\sim$2, due to the 
difference of isotropic indicatrix (line) and dipole indicatrix
(Thomson scattered continuum). The largest equivalent width is
expected for the cloud which is located at the same distance as the
source, but is shifted from the source in the picture plane such that $\sim$90
degree scattered photons are dominating continuum. Variations of
equivalent width due to this effect are clearly see in Fig.\ref{ew} below.

In the light of the above discussion one can rise the question why the
Sgr B2 is brighter in the 6.4 keV line than other
molecular clouds in the region if Srg A* is a source of primary radiation. 
One can consider following possibilities:

\begin{itemize}
\item[\bf I] The parameters of the Sgr B2 cloud (Thomson optical depth, solid
angle occupied by the cloud as seen from the Sgr A* position) may be
favorable for high output in terms of the 6.4 keV line flux. If one
starts with the cloud of very low optical Thomson depth $\tau_T$, then
the flux in the line will be simply 
proportional to $\tau_T$. With increase of $\tau_T$ the
photoabsorption effects start to be important and in the limit of very
thick cloud most of the radiation will be absorbed (unless the side of
the cloud towards observer is illuminated). Optimal for the formation
of the iron fluorescent line is the Thomson depth $\tau_T\sim
0.4$. This value roughly corresponds to the Thomson optical depth of
the Sgr B2 cloud. Thus we conclude that at least 
partially brightness of the Sgr B2 cloud can be attributed to the
`optimal' optical depth of the cloud. If the clouds have
considerable optical depth then their brightness in the 6.4 keV will
strongly depend on what side of the cloud is illuminated (seen by
observer or an opposite one). 
\item[\bf II] The intense primary radiation may be due to a relatively short
episode of efficient accretion onto Sgr A* few hundreds years ago  and
Sgr B2 cloud is located `at the right place' (see Fig.\ref{geom})
such that light propagation 
time from Sgr A* to the cloud and then to the observer compensates the
time delay. In other words only selected clouds (e.g. Sgr B2, Radio Arc
region) are exposed to the primary radiation from Sgr A* at `present'
time. Other clouds in this region either will be bright in future or
primary radiation already left them. 
\item[\bf III] The emission from Sgr A* may be anisotropic, e.g. due
to the blocking by dust at some directions \cite{koy96}, or
intrinsically collimated (e.g. jet) and pointed towards Sgr B2. 
\end{itemize}

Further imaging observations of the Galactic Centre region
in the 6.4 keV line may reveal many smaller clouds, which trace the
volume occupied by the radiation from the putative transient source.

\subsubsection{Long Duration Flare}
In the case of the illumination front broader than the dimension of the
molecular cloud ($c \Delta t > r_{cloud}$) the $K_\alpha$ line surface
brightness is proportional to the optical depth of the cloud (in the
limit of small $\tau_T$) and the total $K_\alpha$ flux is proportional
to the mass of molecular hydrogen inside the cloud (see Sunyaev et al., 1993).

\subsection{Multiple scattering problem}
Even when the primary photons have already left the cloud, multiply
scattered photons can be observed. Indeed, part of the 6.4 keV photons
produced due to absorption of primary continuum radiation will be
Thomson scattered and will leave the cloud with some delay with the
respect to the primary radiation. This delay can be of the order of
light crossing time of the cloud. This is especially important for
clouds located between the compact source and observer, since primary
radiation may leave such clouds on a time scales much shorter than
light crossing time of the cloud. The flux of the radiation leaving
the cloud with significant delay should be small -- for Thomson
depth much less than unity most of the photons leave the cloud without
scattering, while for large Thomson depth the photoabsorption will
significantly attenuate the flux. However, given the present day line
flux from the Sgr B2 cloud (few $10^{-4}~phot~s^{-1} cm^{-2}$, Koyama
et al., 1996 ) it 
is clear that even 100-1000 times weaker flux can be easily detected by future
missions like {\it Constellation} or {\it Xeus}. The time dependence
and the shape of the multiply scattered line is considered in the next
section.

\section{Equivalent Width}
\label{sew}

The case of the source located outside the cloud and illuminating the
side of the cloud towards the observer was intensively discussed in
the literature in application to the spectra reflected by the
surface of accretion disk or dust torus in AGNs (see e.g. Matt et
al. 1991, Awaki et al. 1991, Nandra \& George 1994, 
Ghisellini et al. 1994). We concentrate below on the simple model of a source
embedded into the neutral gas cloud and consider time evolution of
the equivalent width and the shape of the line. 

\begin{figure*}
\plotthree{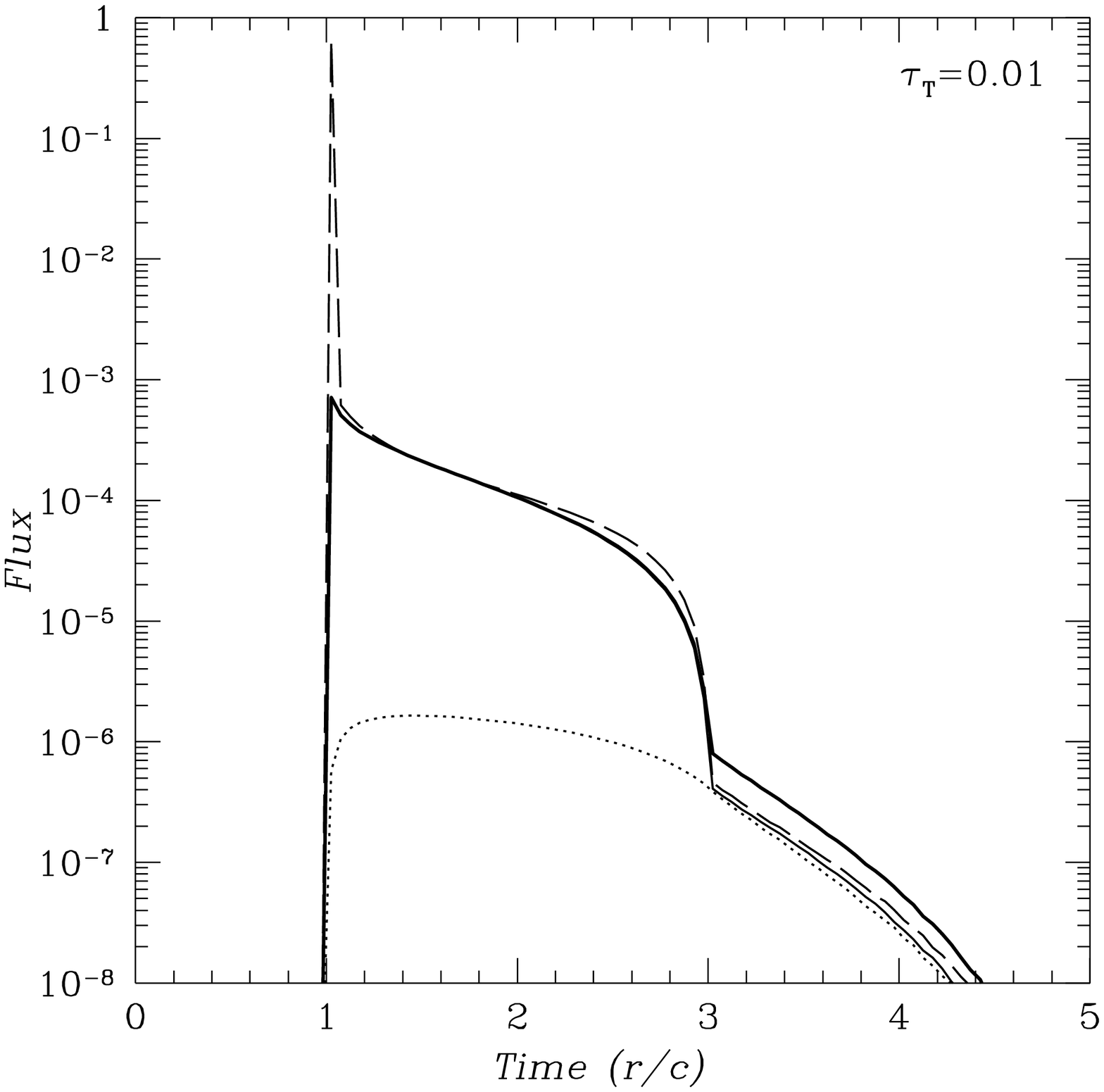}{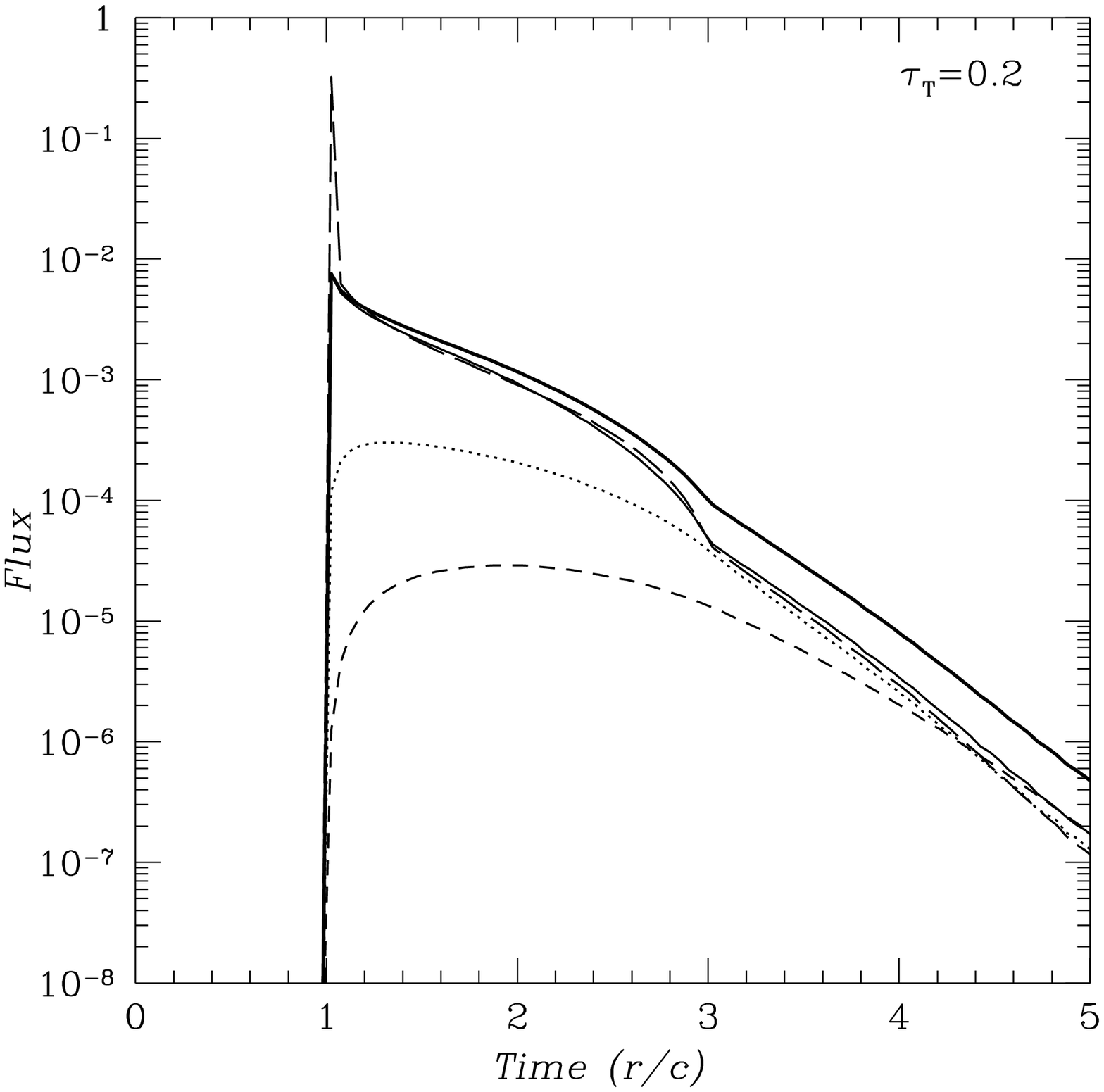}{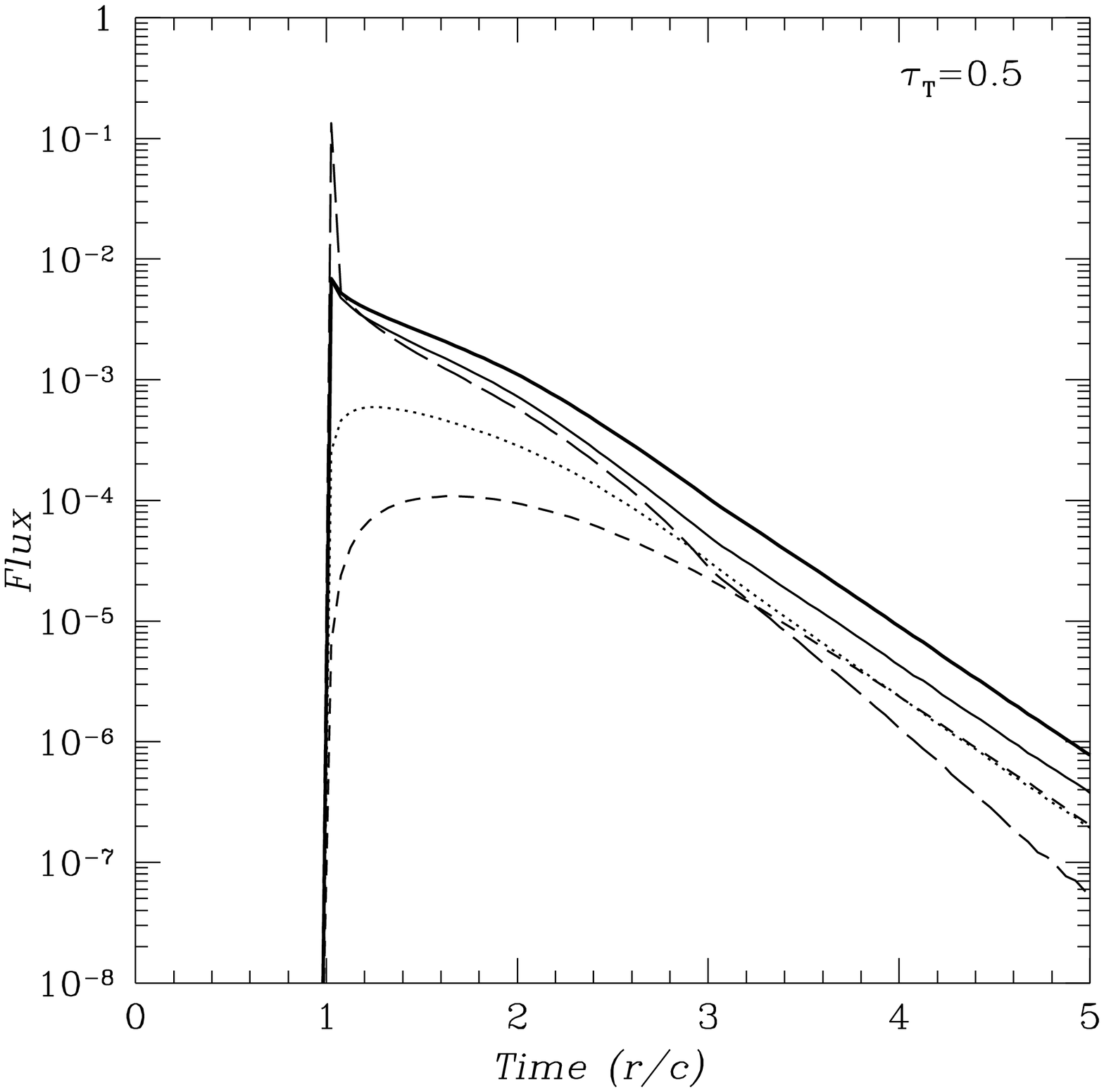}
\plotthree{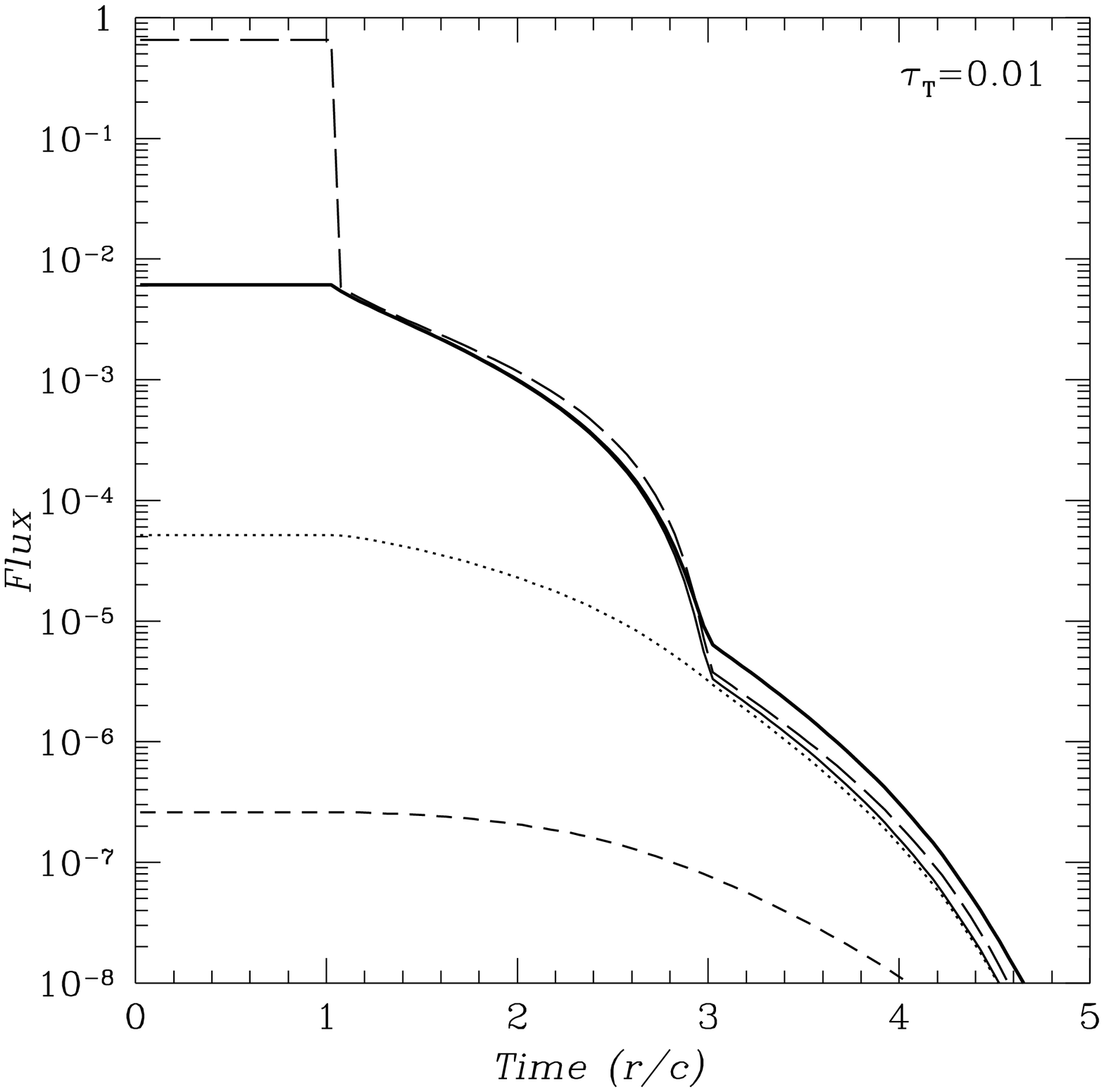}{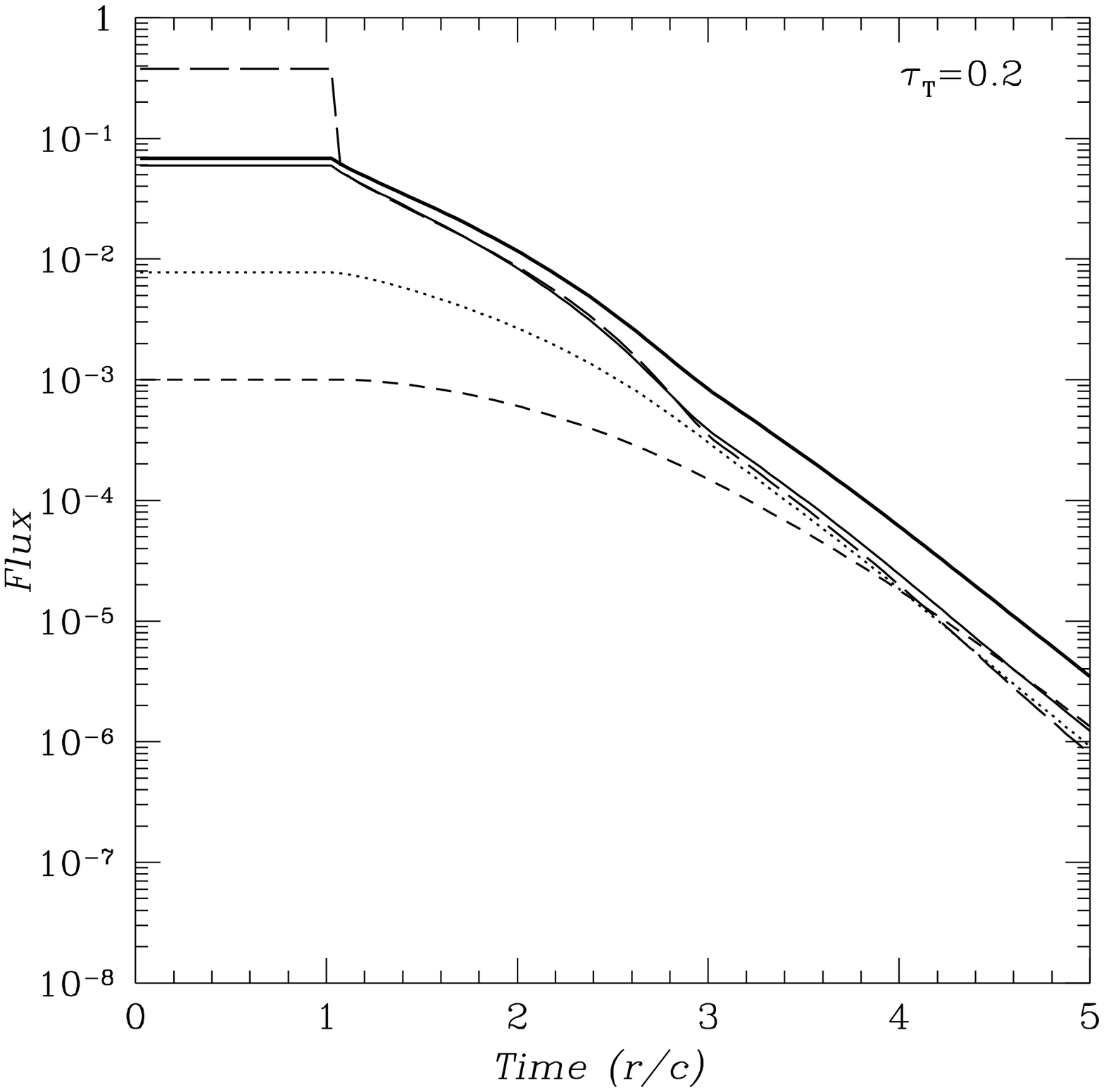}{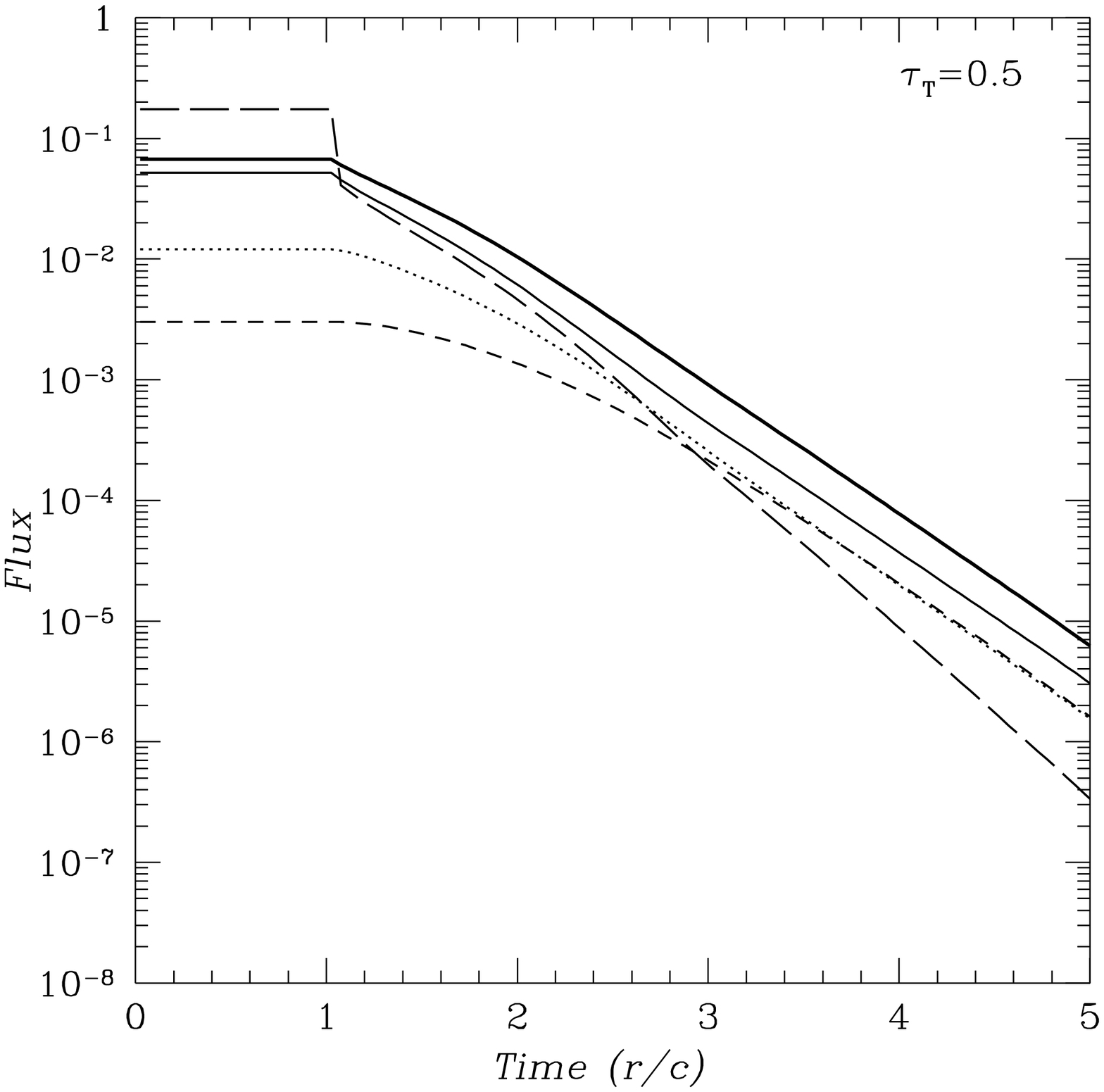}
\caption{Flux versus time for the compact source of continuum emission at
the centre of the uniform spherical cloud. The upper row of figures corresponds
to the short flare of the compact source flux, the lower row of figures
corresponds to the `switch--off' of the steady source. In both cases
the propagation time from the centre of the cloud to the outer edge
was taken into account. Thus flare at the centre of the cloud at $t=0$
appears as a spike at $t/(r/c)=1$ in the figures. The thin 
solid line shows the contribution of the 6.4 keV photons, which left the
cloud without further interaction. The dotted line shows the contribution of
6.4 keV photons which have been Thomson scattered once and dashed line
shows the 6.4 keV photons which undergo more than one scattering before
escaping from the cloud. The thick solid line shows total contribution of
all these components. At last long--dash line shows the continuum flux (in
the energy range [5.9--6.9] keV). Time is expressed in units of light
crossing time of the cloud (radius divided by speed of light). 
} 
\label{flux}
\end{figure*}

\begin{figure*}
\plotthree{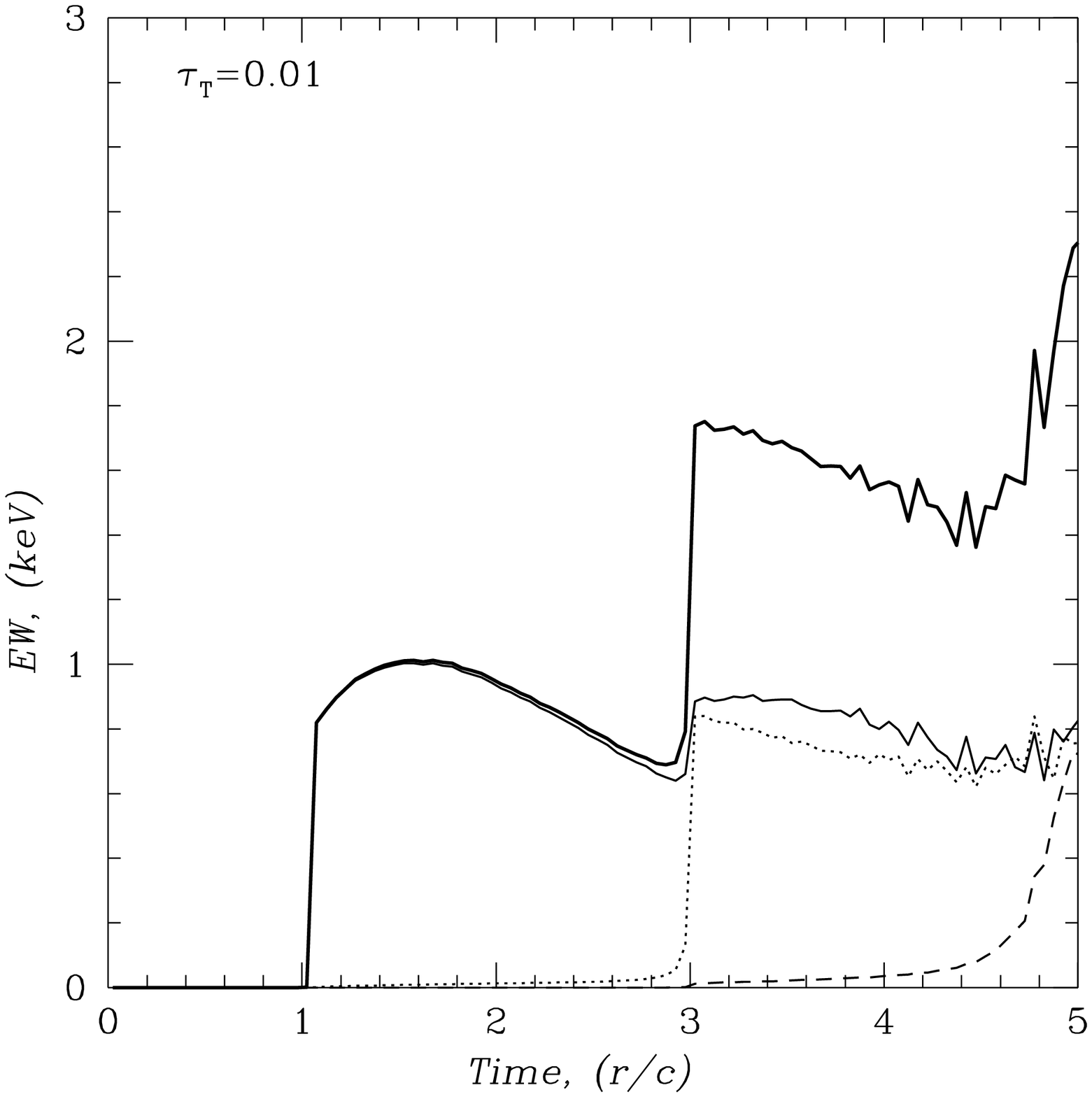}{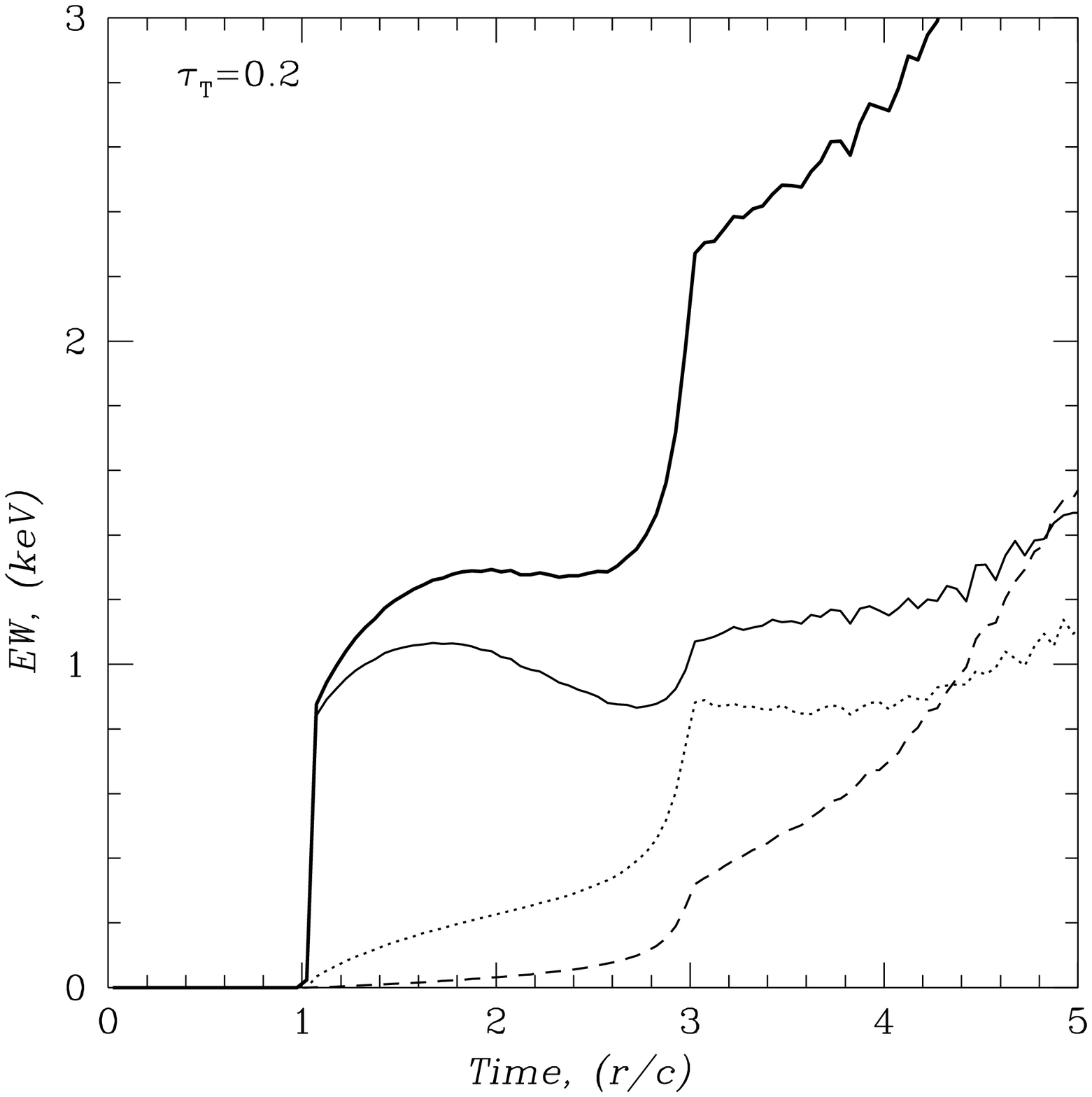}{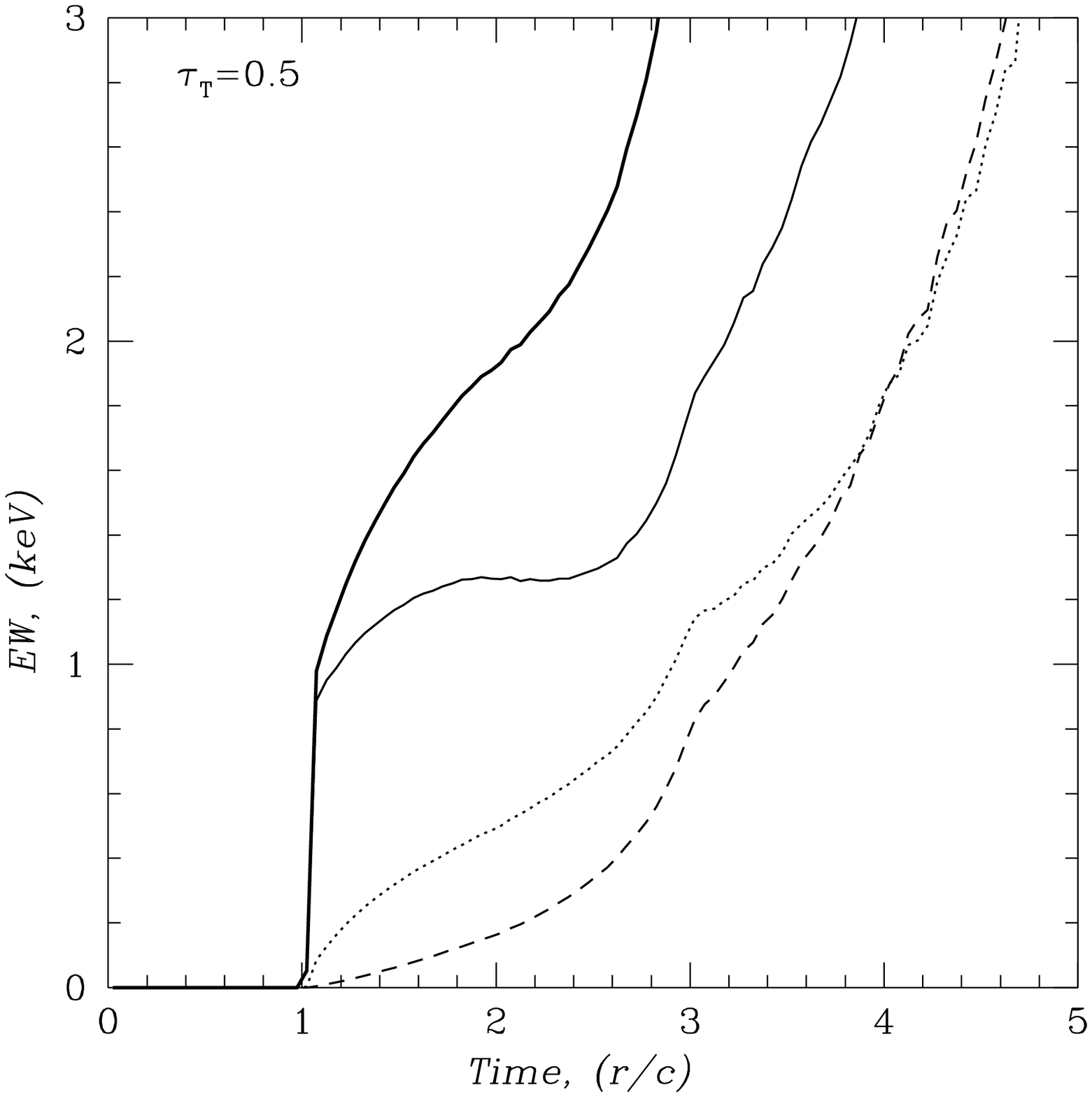}
\plotthree{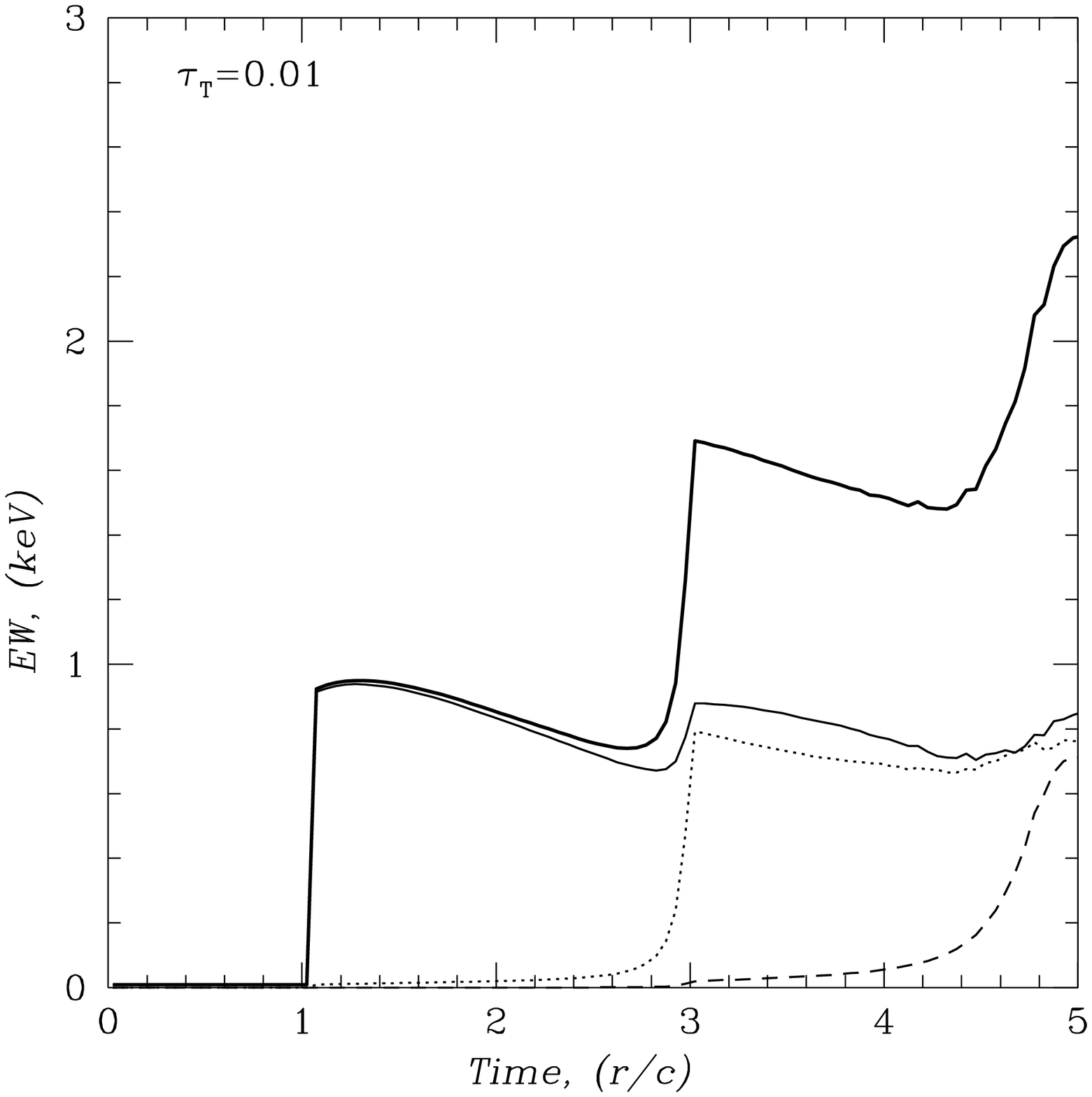}{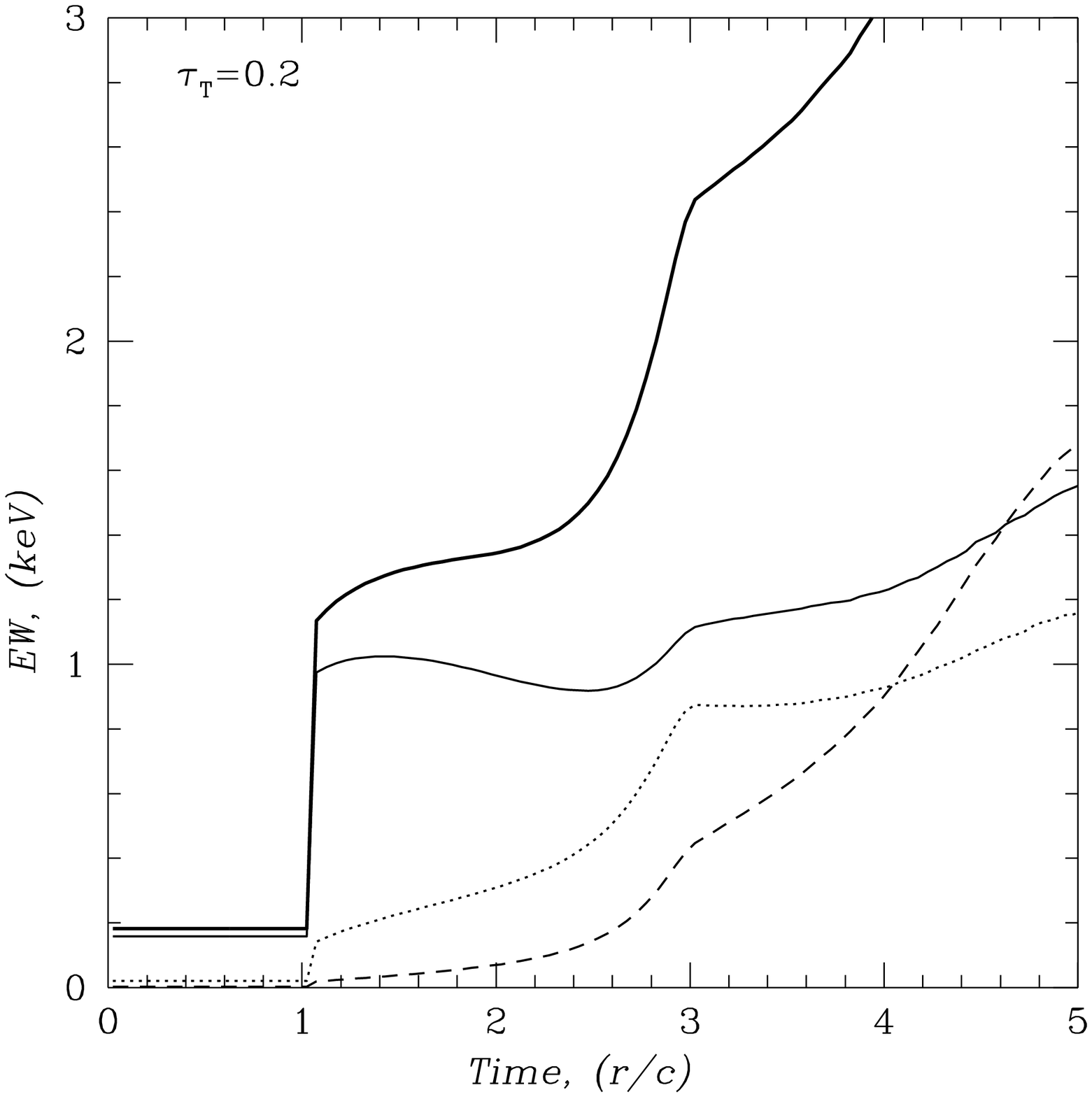}{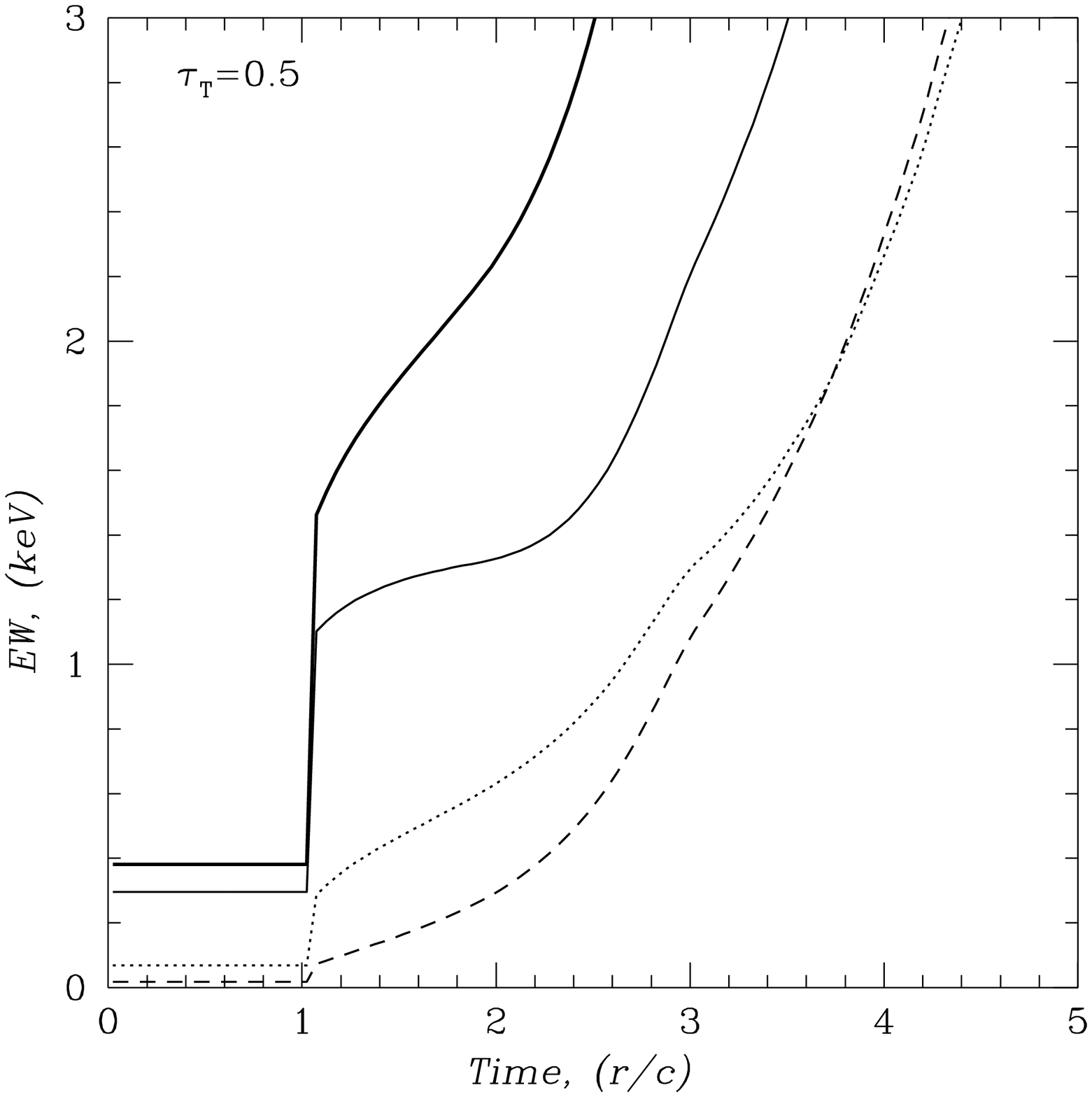}
\caption{
The equivalent width versus time for the compact source of continuum
emission at 
the centre of the uniform spherical cloud. The upper row of figures corresponds
to the short flare of the compact source flux, the lower row of figures
corresponds to the `switch--off' of the steady source. The thin 
solid line shows the contribution of the 6.4 keV photons, which left the
cloud without further interactions. The dotted line shows the contribution of
6.4 keV photons which have been Thomson scattered once and dashed line
shows the 6.4 keV photons which undergo more than one scattering before
escape from the cloud. The thick solid line shows total contribution of
all these components. 
} 
\label{ew}
\end{figure*}

        One of the most important parameters determining the possibility of
the line detection is its equivalent width (EW), i.e. the ratio of the flux in
the line to the spectral density of the continuum flux at the same energy.
We use Monte--Carlo simulations in order to generate spectra, emerging from a
spherically symmetric cloud with uniform density and a point source of
the continuum radiation 
(power law with photon index $\Gamma=1.8$) in the centre. Initial seed
photons were generated in the 1--18 keV energy range. Since temporal
behavior of the 6.4 keV line flux is in question we kept record of escape
time for photons, emerging from the cloud. Shown in Fig.\ref{flux} is the
evolution of the line and continuum fluxes with time for the case of short
flare at $t=0$ (upper row of figures) and theta--function--like  behavior of
primary source flux (lower row of figures). In the latter case the flux of the
source was stable before the moment of time $t=0$ and dropped to zero
level after this moment. In both cases  $t=1$ corresponds to the
moment of time, 
when direct radiation (towards observer) leaves the cloud.  
Three values of the cloud radial Thomson depths $\tau_T=$ 0.01 (a), 0.2 (b)
and 0.5 (c) were used. In Fig.\ref{ew} the evolution of the  equivalent
width of the 6.4 keV line is shown for the same set of parameters. 
The thin 
solid line shows the contribution of the 6.4 keV photons, which left the
cloud without further interactions. The dotted line shows the contribution of
the 6.4 keV photons which have been Thomson scattered once and the dashed line
shows the 6.4 keV photons which undergo more than one scattering before they
escaped from the cloud. At last thick solid line shows total contribution of
all these components. Note that scattered 6.4 keV photons may have an actual
energy lower than 6.4 keV (we discuss the shape of the line profile below).

The first thing apparent from Fig.\ref{ew} is the sharp change of EW at the
$t=1$. This is 
obviously related to the disappearance of the direct component (escaping
from the 
cloud without interaction), which contributes to the continuum emission.
Immediately after $t=1$ EW jumps to the level of $\sim$ 1 keV for all three
values of $\tau_T$ (Fig.\ref{ew} a,b,c). Indeed (see e.g. Vainshtein
\& Sunyaev, 1980) 
both flux in the line and  
the level of the scattered continuum emission are (to the first order)
proportional to the amount of matter over the radius of the cloud. As the
results, the EW (ratio of the line flux to the continuum spectral density)
happens to depend only weakly on the actual parameters of the cloud.

Note also (see Fig.\ref{ew}a) the specific (curved) shape of the EW behavior
in the  
period of time $t=[1,3]$, when single scattered continuum photons and
unscattered 6.4 keV photons dominate the spectrum. This shape (a deviation
from the constant 
level) is due to the difference in the angular distribution of the
scattered continuum photons and those in the 6.4 keV line. Indeed 6.4 keV
photons are emitted by iron atoms isotropically, while continuum photons,
scattered by electrons, follow Rayleigh (dipole) indicatrix
$\frac{d\sigma}{d\Omega}\propto (1+cos^2\vartheta)$. A difference in the
escape time for continuum and line photons causes this specific shape.

\begin{figure}
\plotone{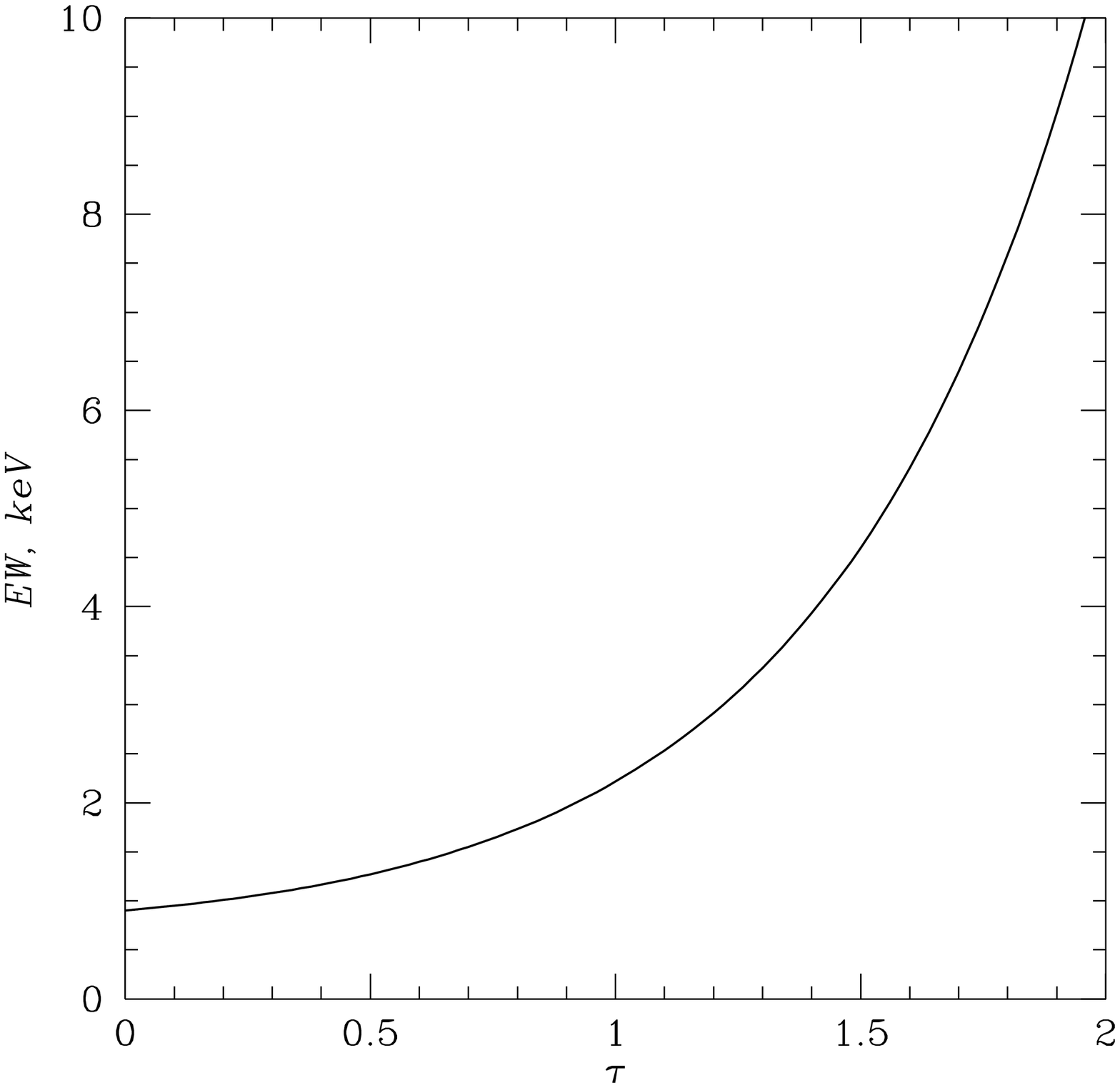}
\caption
{Dependence of the equivalent width on the optical depth of the media. 
The photons at $\sim$ 6 keV have a higher probability as being
absorbed, compared to high energy photons (10--20 keV), which can pass
through larger depth  and cause emission of the 6.4 keV photons. One
can expect to find such an equivalent width in the scattered
component if the radiation has passed through the absorbing matter
with the optical depth $\tau_T$ before entering the region from which
reprocess radiation is observed.} 
\label{ewtau}
\end{figure}

However for large values of optical depth (Fig.\ref{ew}c) the behavior of
the EW 
is very different -- its value starts to rise from early
beginning. One of the reasons for such a behavior is a significant
photoabsorption depth. Continuum 
photons (at about 6 keV) have a high probability to be absorbed 
before they can escape from the cloud. On the other hand high energy photons
(at the energy 10--20--30 keV) can reach peripheral regions of the cloud
without being absorbed, ionize an electron from the iron K--shell and thus
produce a 6.4 keV photon, which can escape further. One can get an idea of
how the EW depends on the depth of the scattering region from
Fig.\ref{ewtau}. The following 
function $\frac{Y\cdot\delta_{Fe}\int^\infty_{7.1 keV} I(E) 
exp(-\sigma_{ph}(E)\tau_T/\sigma_T)\sigma_{Fe}(E)dE }{I(E_0)
exp(-\sigma_{ph}(E_0)\tau_T/\sigma_T)}$ is plotted here versus the optical
depth  $\tau_T$. This function  
characterizes how the equivalent width of the line in the scattered
component changes if the primary radiation has passed though the
absorbing matter before the scattering place. Absorption just modifies
the shape of the spectrum, suppressing continuum at $\sim$6 keV
stronger than at energies above $\sim$ 10 keV. As a result one can
expect that for large values of 
the optical depth the EW of the line in the scattered component will
be larger than that in the case of small $\tau_T$.  

 Even apart from the photoabsorption effect the total EW calculated as the
sum of 
unscattered and scattered 6.4 keV photons (Fig.\ref{ew} a,b,c) grows with
time. 
For low values of optical depth this growth is seen as the sharp jump at
$t=3 R/c$. This jump corresponds to the moment when all photons, which
undergo only one interaction (absorption $+$ fluorescence or single Thomson
scattering) are leaving the cloud. After that moment of time the spectrum is
dominated by the photons passed through more than one interaction. One can
see that the EW 
of the 6.4 keV complex increases at that moment. This is because 
continuum is now dominated by twice Thomson scattered photons, while the
line is 
constructed from two kinds of photons: (a) photons which first were Thomson
scattered, then ionized an electron from iron K--shell and produced new 6.4
keV photon and (b) Thomson scattered 6.4 keV photons. The EW for each of
these components (with respect to twice Thomson scattered continuum
photons) is about 1 keV. After each additional scattering (when
the spectrum is dominated by the photons passed through $n$ interactions)
the total equivalent width will increase further (roughly proportional to
$n$). In other words one can say that with each additional `interaction'
Thomson scattering does not
change the existing EW of the line, while ionization of iron K-shells adds more
`fresh' 6.4 keV photons.   

Finally one can note that although the EW of the 6.4 keV complex rises
with time absolute flux declines with time (Fig.\ref{flux}). The decline of
the flux is caused by the escape of the photons from the cloud and
photoabsorption of photons. The fraction of photons which can be
detected after additional light crossing time of the cloud is less
than $\sim\tau_T e^{-\tau_T \sigma_{ph}(E)/\sigma_T}$. This factor
reaches maximum ($\sim 0.14$) at Thomson depth of the cloud
$\tau_T\sim 0.4$. Note that for the Sgr B2 
complex ASCA detected the 6.4 keV flux at the 
level of few $10^{-4} phot~s^{-1} cm^{-2}$ \cite{koy96} and even a 1000 times
weaker line will still give a few hundred counts for 100 ksec
observation with the {\it Constellation} mission. It means that in principle
with {\it Constellation} it may be possible
to detect traces of the 6.4 keV emission even if the time elapsed
since the initial flare exceeds the light crossing time of the cloud
by a factor of few. In order to multiple scattered photons dominate
the spectrum the cloud has to have rather sharp boundary. As noted
above although the life time of the $n$--times scattered photons is
longer than for photons which undergo $n-1$ scatterings, the flux
decreases by a factor of at least $0.14$ (and in fact more, see
Fig.\ref{flux}) with each additional scattering. If a very extended
envelope (with weakly declining density) 
of the cloud is present or if the primary continuum flux was declining
too slowly, then fluorescent photons generated by primary radiation
may dominate the total spectrum of the cloud. However in the presence
of optically thick blobs immersed into more tenuous phase they may
effectively screen more distant layers of the cloud (which are still
exposed to primary radiation) while their surface seen by the observer
will reflect 6.4 keV photons. 

\section{Shape of the Line}
\label{sshape}
In the previous section the fact that photons can change their energy during
scattering was not taken into account. However for the 6.4 keV photons even
single scattering causes significant change of energy, which is especially
important for future instruments with high energy resolution. Since we are
considering scattering by the neutral matter, then all complications related
to bound electrons (discussed in Sunyaev \& Churazov, 1996) are to be taken
into account. Although we are considering molecular cloud, for
simplicity we used below 
analytical formulae for scattering by hydrogen atoms. Compton
scattering (affected by momentum distribution of the electron) should
be very similar for atomic and molecular hydrogen. 

For the 6.4 keV photons after each scattering
(by a hydrogen atom) about 14 per cent of the photons will maintain the initial energy
unchanged (Rayleigh or elastic scattering). For molecular hydrogen
this value is higher due to the coherent scattering by two electrons. 
Some small fraction photons will cause excitation of electrons in the
hydrogen atoms, leading to the appearance of Raman satellites of the
line at the energies $h\nu_0-13.6\times (1-1/n^2)$~eV, where $n$ is
the principle quantum number of excited level. But the largest
fraction of photons will be Compton scattered by 
hydrogen atoms, leading to the ionization of an electron and decrease of the
photon energy by $\sim$ 13--200~eV. Due to the motion of the electrons bound in
hydrogen atoms, the  backscattering peak will be smeared out. 
\begin{figure*}
\plottwo{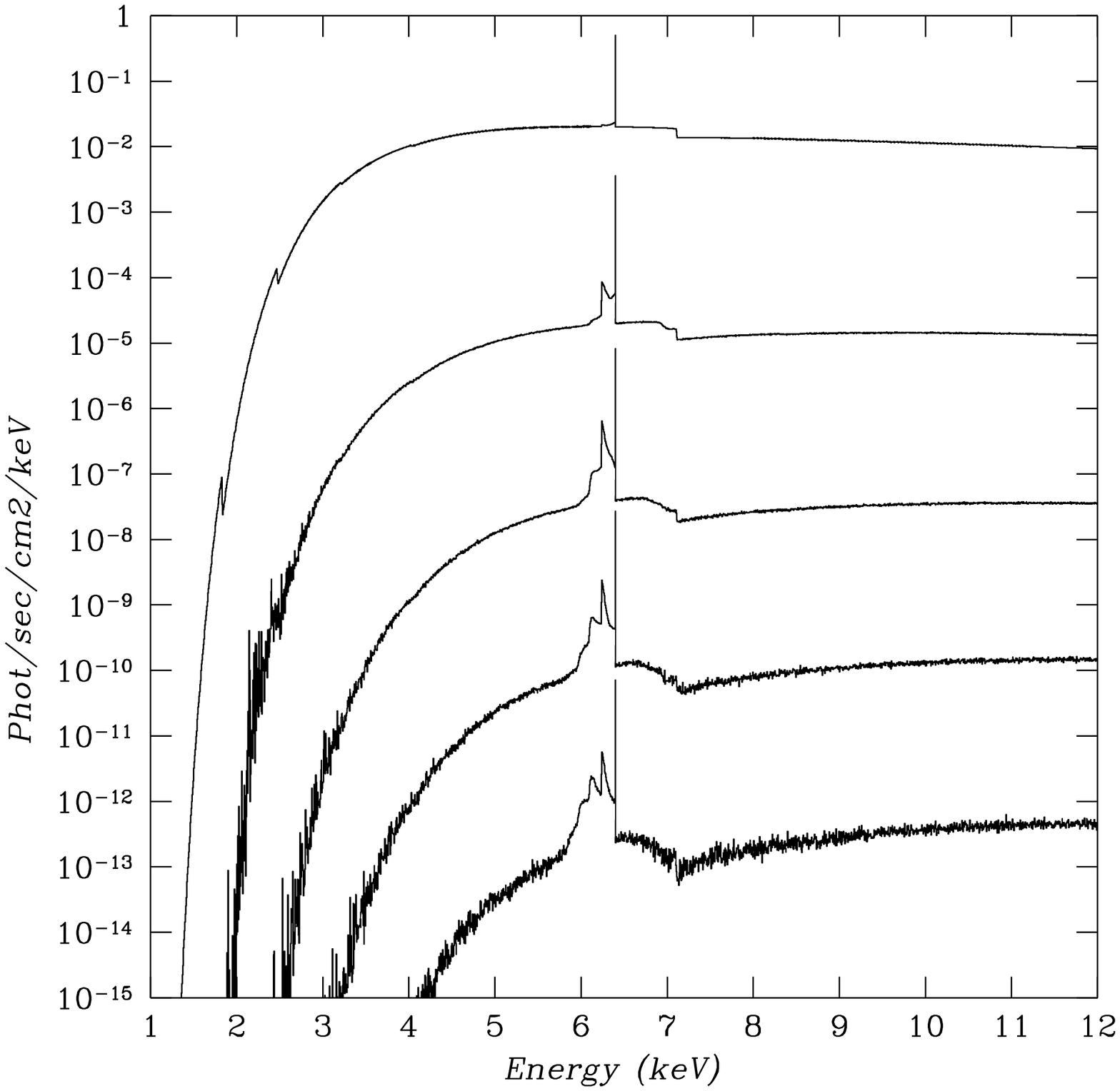}{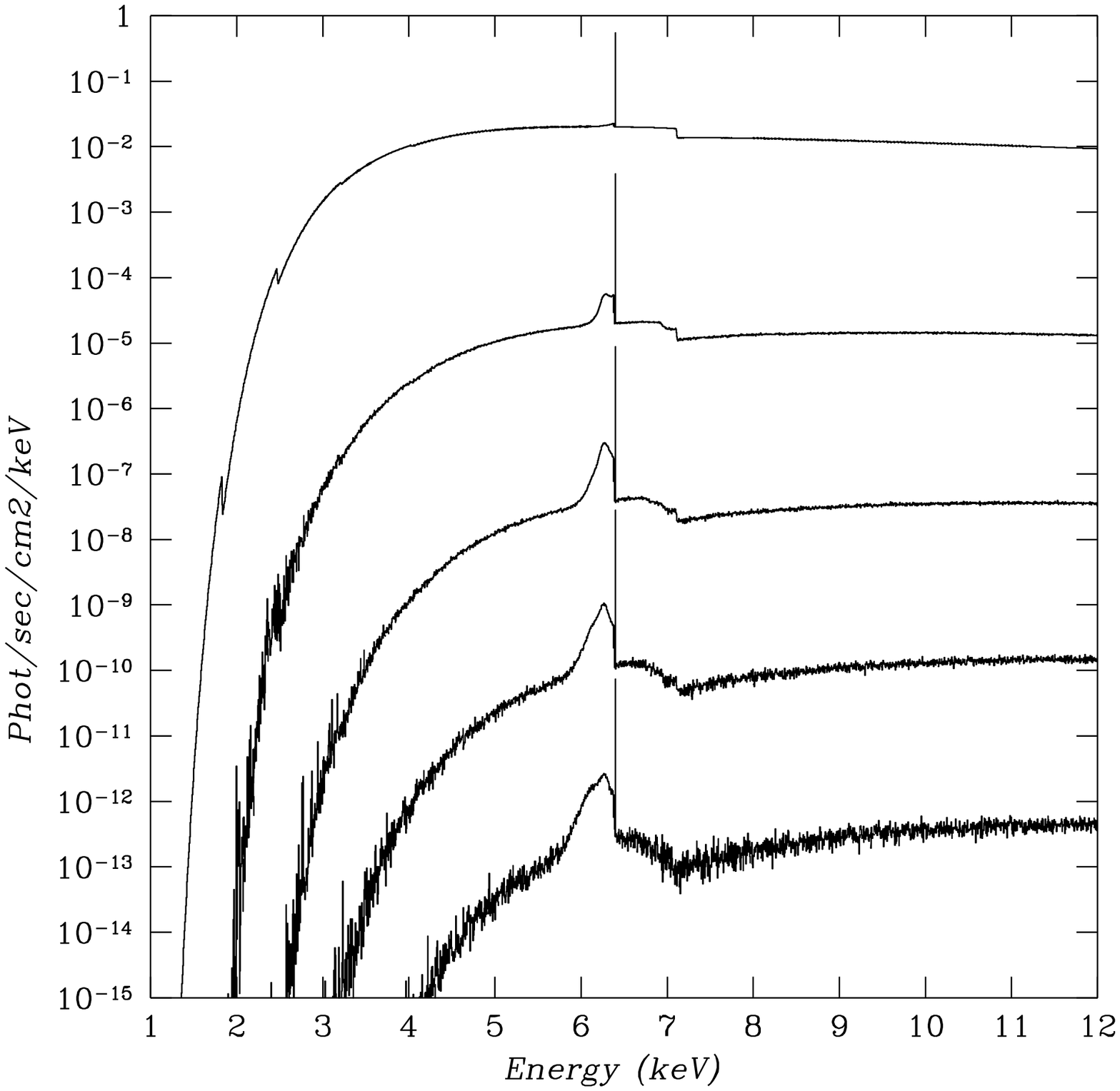}
\caption{
Spectra emerging from the uniform cloud with radial Thomson depth
$\tau_T=0.2$ at different moments of time after a short flare of the continuum
emission from the point source in the centre of the cloud. For the left
figure the recoil effect was calculated as for free cold electrons at rest. For
the right figure effects of bound electrons were taken into account (but only
for the 6.4 keV line). Spectra correspond to the intervals of time: 0--1,
1--2, 2--3, 3--4, 4--5 (in units $R/c$). Each subsequent spectrum was
multiplied by 0.05 for clarity. Spectra are plotted with an energy resolution
of 5 eV.} 
\label{tspec}
\end{figure*}

\begin{figure}
\plotone{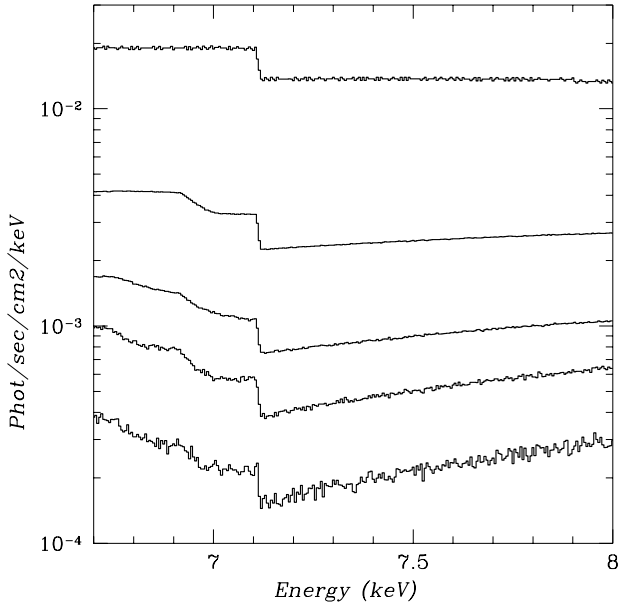}
\caption{
Evolution of the iron absorption K--edge with time. Spectra correspond
to the intervals of time: 0--1, 1--2, 2--3, 3--4, 4--5 (in units
$R/c$). Each subsequent spectrum was multiplied by a factor of 10.} 
\label{edge}
\end{figure}

Shown in Fig.\ref{tspec} are the spectra, emerging from the spherical cloud
(with the radial optical depths of $0.2$) after a
given time from the switch--off of the compact source. We assume here that
fluorescence produces a monochromatic line at the energy of 6.4 keV. One can
see that after multiple scatterings a complex shoulder forms on the low energy
side of the line. As discussed above the total equivalent width of the whole
complex increases with time. However the photons are now distributed
over a rather broad energy range, which increases with time. 
The equivalent width of the unshifted 6.4 keV line itself will still
increase with time. This is related in part with the more rapid decline of the
continuum at 6.4 keV due to photoabsorption (see above)
and in part with Rayleigh scattering, which leaves $\sim$14 per cent
of the photons at the initial energy (see also below).

\section{Absorption edge}
High sensitivity and energy resolution of the coming missions will be
sufficient to study the absorption edge (at 7.1 keV) in much more
details. Shown in Fig.\ref{edge} is the evolution of the absorption
edge structure for the similar sets of moments of time as in
Fig.\ref{tspec}. Since the feature is intrinsically broad we ignored
the complications associated with scatterings by bound electrons and used
standard formulae for recoil effect on the free electrons in order to
study evolution of the shape of the spectrum.  Multiple
scattering cause shift of the feature to lower energies and smearing
of the edge structure. Sharp jump at 7.1 keV has
always the same relative amplitude, which simply reflects the
absorption depth after the last scattering. 

For the case of a steady source inside a cloud the observed flux in the 6.4 keV
line is uniquely related to the observed deficit of photons above
K-edge. The 
coefficient relating these two quantities is simply the fluorescent
yield $Y\approx 0.3$.
In other words equivalent width of the line is approximately factor of
3 smaller than effective equivalent width of the absorption edge.
For the reprocessed radiation (i.e. when primary source is not
visible) this relation may break. This is clear considering e.g. the case
of the cloud of extremely low optical depth. Indeed as one can see
from Fig.\ref{ew} (left column) equivalent width emerging from an
optically thin cloud can be high. On the other
hand observed deficit around 7.1 keV (even if we integrate it with
account of smeared absorption structure) can be very small for low
values of optical depth of the cloud.

\section{Complex Geometries}
\label{scomplex}
Of course in the case of more complex geometries (e.g. extended
envelopes) and complex time history of 
the primary radiation flux the whole picture will be accordingly much more
complex. 
However some general properties (which are demonstrated with
the above simple models) may be used to get an idea of the what we can
look for with future X-ray observations of the Galactic Centre region
in order to understand the origin of the fluorescent line. 

The high sensitivity and high angular resolution mapping of this
region in the 6.4 keV may reveal a number of small clouds (along with
more detailed structure of the Sgr B2 cloud and Radio Arc region)
which are tracing the volume occupied by primary radiation of putative
flare of the Sgr A*. Morphology of the line surface brightness
distribution may help to determine the location of the primary
source. Broad band spectra of these clouds may 
also provide an information on the primary source location, since low
energy absorption depends crucially on the side from which the cloud
is illuminated. Indeed, if the side of the cloud towards observer is
illuminated by a primary source, then instead of exponential cutoff at
low energies due to photoabsorption (as in the case of the cloud
illuminated from the back) one can expect much shallower decline.

For some clouds inspite of the low absolute flux the 
equivalent width of the line can be
in excess of 1 keV   due to photoabsorption effects or multiple
scatterings\footnote{Of course, the extended emission due to
the hot gas in the Galactic Centre region may actually dominate the
continuum at 6.4 keV causing the decrease of the observed equivalent
width. However one can estimate it's contribution using the
intensities of the strongly ionized iron emission lines, in
particularly at $\sim$ 6.7 keV.}. Narrow line and sharp absorption
edge at 7.1 keV can be considered 
as the indicators of strong photoabsorption, while the complex shape of the
line (extended left wing) and smeared absorption edge indicate that
multiple scatterings are dominant. 

Important results can be obtained studying dense condensations which
are optically thick. Since they screen the more distant layers of the
gas one can use them as indicators of the position of surface which
separates regions filled with primary photons from `empty' regions.
From such condensations scattered 6.4 keV line may be detected even if
the cloud has extended envelope.

Principle information can be obtained comparing the data 
which were obtained at different moments of time (separated
by $\sim 5$ years). Detection of changes in the surface brightness, equivalent
width and shape of the 6.4 keV would provide exciting and unique
insight on the time history of the putative supermassive black hole at
the Centre of our Galaxy.

\section{Conclusions}
\label{ssum}
The value of the equivalent width and the shape of the 6.4 keV iron
$K_\alpha$ line, emerging from the neutral matter, illuminated from inside
or outside by the X--ray continuum spectrum, contains information on the time
history of the illuminating continuum flux. Assuming normal abundance of
iron in the scattering media the value of an equivalent width of
about 1 keV indicate that we are dealing with a scattered component. Values in
excess of 1 keV can be due to strong photoabsorption or multiple Thomson
scattering. Spectroscopic analysis of the 6.4 keV line profile and the shape
of the 7.1 keV absorption edge can help to distinguish between these
possibilities. When combined with broad band
spectroscopic measurements it can be used to determine the position and flux
history of the primary continuum source. A new generation of X--ray
instruments should be capable of achieving the required sensitivity and energy
resolution to accurately measure the detailed structure of the 6.4 keV line in
the Galactic Centre region. Comparing the flux and shape of the line from
different molecular clouds in the region, one can reconstruct the date and
duration of the flare, responsible for observed reprocessed emission. Note
also that observations, over a period of 5--10
years, may show the variability of the line flux, shape and morphology
of it's surface brightness distribution.

The assumption that Sgr A$^*$ emission has caused the 6.4 keV line (and
continuum) emission from Sgr B2 \cite{smp93,koy96} 
implies that its flux vary by a factor of at least $10^3$ (and perhaps much
more) on a time scale of hundreds of years. If nuclei in other galaxies
also have similar behavior there is a good chance of detecting `delayed'
scattered components from some of them. With the high sensitivity of
the {\it Constellation} mission 
one can search for very weak nearby and distant AGNs. The detection of
an iron line of large equivalent width and complex shape would prove
the past violent activity of the central engines of these sources.

This work was supported in part by the grants RBRF 96-02-18588 and
INTAS 93-3364-ext. We thank Marat Gilfanov for useful discussions and
anonymous referee for helpful comments.

\appendix

\section{Simplified Treatment}
\label{ssimlpe}
	An exact account for complex geometry, scattering diagram
and photoabsorption effects make the Monte--Carlo method most suitable for
solving each particular problem. However all major effects can be easily
seen from the simplified treatment, given below.
\subsection{Equivalent Width}
	Let us consider the evolution of spectrum for the short flare of
continuum emission in a cloud of Thomson (and photoabsorption) depth
$\tau_T \ll 1$. One can construct a `toy' model assuming that fraction
of photons $\sim \tau_T$, which undergo additional scattering, will
leave the cloud with time delay $T$. Therefore intensity of the
continuum (at 6.4 keV and above iron 
K--edge) at a given moment of time $t$ can be written as follows:
\begin{eqnarray}        \label{nc6}
N_c^{6.4}(t)=N_c^{6.4}(t-T)\cdot \tau_T
\end{eqnarray} 
\begin{eqnarray}        \label{nc7}
N_c^{7.1}(t)=N_c^{7.1}(t-T)\cdot \tau_T
\end{eqnarray} 
We further assume that similar delay $T$ is applicable to the
continuum photons, which have been absorbed by iron K shell and
reemitted as 6.4 keV photons. Therefore:
\begin{eqnarray}        \label{nl}
N_l^{6.4}(t)=N_l^{6.4}(t-T)\cdot \tau_T+N_c^{7.1}(t-T)\cdot \tau_T Y
\delta_{Fe} \frac{\sigma_{Fe}}{\sigma_T}
\end{eqnarray} 
First term in the right side of equation (\ref{nl}) corresponds to
scattering of existing 6.4 keV photons, while the second term
describes generation of new 6.4 keV photons. Dividing equation
(\ref{nl}) by the equation (\ref{nc6}) one gets an expression for the
equivalent width of the line:
\begin{eqnarray}        \label{new}
EW(t)=EW(t-T)+\frac{N_c^{7.1}(t-T)}{N_c^{6.4}(t-T)}\cdot Y
\delta_{Fe} \frac{\sigma_{Fe}}{\sigma_T}
\end{eqnarray} 
The last term in the equation (\ref{new}) does not depend on time or
Thomson depth and it's numerical value is $\sim 1$ keV (see previous
sections). Thus from equation (\ref{new}) it is clear that equivalent
width rises by $\sim 1$ keV with each additional time interval $T$.
\begin{eqnarray}        \label{new1}
EW(t)=EW(t-T)+ 1~keV
\end{eqnarray} 
Account for photoabsorption causes additional increase of the
equivalent width with time.

\subsection{Shape of the line}
\begin{figure}
\plotone{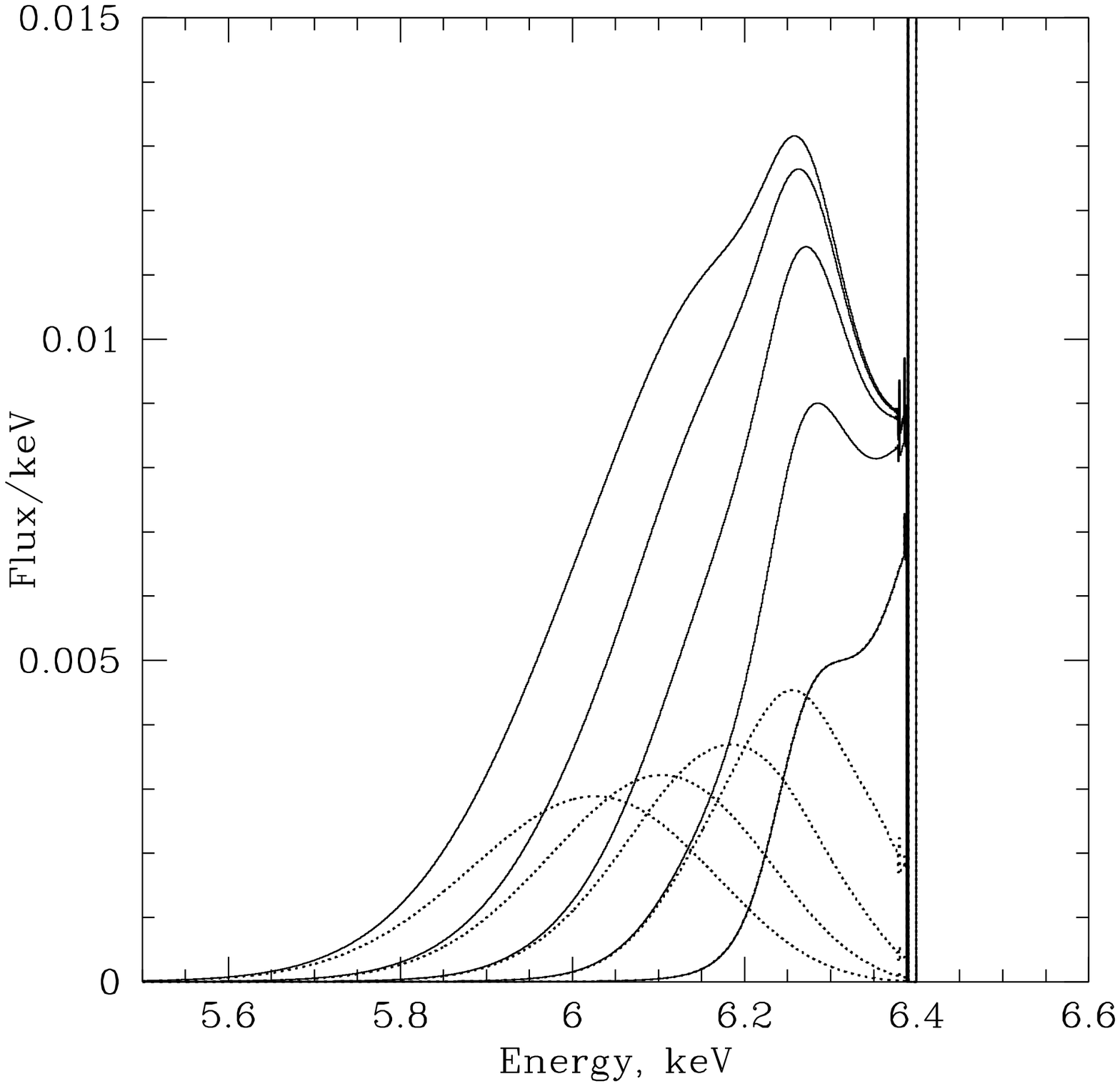}
\caption
{Shape of the line after the period of time $\frac{t}{R/c}=1,2,3,4,5$ since
short flare of continuum emission (simplified treatment, see
text). The dotted lines show the spectra of a monochromatic line after
1, 2, 3, 4, 5 scatterings.
} 
\label{shape_a}
\end{figure}
	Considering again the case of $\tau_T \ll 1$ and neglecting the
correlation of the escape probability with scattering angle  one can
calculate the spectrum of the 6.4 keV line complex 
after $t=n R/c$. This spectrum can be represented as the sum of spectra
$\sum_1^n F_n(E)$, each having the same flux. $F_1(E)$ is the `fresh' 6.4
keV line, which will be treated below as monochromatic, each $F_{i+1}$ can
be derived from $F_{i}$, convolving $F_{i}$ with the differential
cross section of scattering, averaged over all angles. Using the differential
cross section for a hydrogen atom (e.g. Sunyaev \& Churazov, 1996) one can get
the sequence of spectra shown in Fig.\ref{shape_a} (cf. Fig.1 in
Illarionov et al., 1979 for the case of scatterings on the cold free
electrons). Dotted lines show individual $F_i$ spectra, while solid 
lines show sums of $F_i$ for $n=1,2,3,4,5$. As noted above, the total flux
of the  line will strongly decline with time, but high equivalent
width will, on the 
contrary, rise, which may help to detect and resolve the shape of the line
even if the absolute flux is low. One can see that the shape of the feature
closely resembles the one obtained from Monte--Carlo calculations
(Fig.\ref{tspec}), although quantitatively they may differ (e.g. because of the
correlation of the escape probability with scattering angle). In particular
this correlation may be important for the contribution  to the intensity of the
unshifted 6.4 keV line due to Rayleigh scattering, which corresponds to
small scattering angles. In the simplified treatment (as above) the width of
the line after $n$ scatterings is equal to $\approx 1+(0.14)+...(0.14)^n$,
where the factor $\sim 0.14$ is the probability of Rayleigh scattering for
the $\sim 6.4$ keV photons, averaged over all scattering angles. 
In reality the intensity of this line will be lower, since photons,
scattered by small angle continue to move towards edges of scattering
media. For molecular hydrogen (which is of course the dominant part of the
cloud) due to coherent scattering the fraction of elastic scatterings
(as well as total scattering cross section) will be higher. Helium
atoms also contribute significantly to the coherent
scattering (Vainshtein, Sunyaev \& Churazov, 1998). However elastic
scattering dominates at rather small 
scattering angles thus reducing the impact of enhanced elastic
scattering onto observed spectra. Monte--Carlo simulations of the
particular geometry of the scattering cloud are required to calculate
the actual fraction of unshifted 6.4 keV line in the observed spectrum.

\label{lastpage}


\begin{thebibliography}{}
\bibitem[\protect\citename{Awaki et al.\ }1991]{awa91} Awaki H.,
        Koyama K., Inoue H., Halpern J.P.,  1991, \pasj, 43, 195
\bibitem[\protect\citename{Bambinek et al.\ }1972]{bam72} Bambinek
        W. et al. 1972, Rev. of Modern. Phys., 44, 716
\bibitem[\protect\citename{Basko }1978]{bas78} Basko M. 1978, \apj, 223, 268 
\bibitem[\protect\citename{Basko et al.\ }1974]{bst74} Basko M.,
        Sunyaev R., Titarchuk L., 1974, \aap, 31, 249  
\bibitem[\protect\citename{Beckert et al.\ }1996]{bdm96} Beckert T.,
        Duschl W.J., Mezger P.G., Zylka R., 1996, \aap, 307, 450  
\bibitem[\protect\citename{Chevalier }1986]{che86} Chevalier R., 1986,
        \apj, 308, 225 
\bibitem[\protect\citename{Churazov, Gilfanov, Sunyaev }1996]{chu96}
         Churazov E., Gilfanov M., Sunyaev R., 1996, \apj, 464, L71 
\bibitem[\protect\citename{Fabian }1977]{fab77} Fabian A.C., 1977,
         \nat, 269, 672
\bibitem[\protect\citename{Hasegawa et al.\ }1994]{has94} Hasegawa T.,
         Sato F., Whiteoak J., Miyawaki R., 1994, \apj, 429, L77 
\bibitem[\protect\citename{George \& Fabian }1991]{gf91} George
         I.M. \& Fabian A.C., 1991, \mnras, 249, 352
\bibitem[\protect\citename{Genzel et al.\ }1994]{gen94} Genzel R.,
         Hollenbach D., Townes C., 1994, Rep. Prog. Phys., 57, 417
\bibitem[\protect\citename{Gilfanov, Sunyaev, Churazov }1987]{gsc87}
         Gilfanov M., Sunyaev R., Churazov E., 1987, 
         Soviet~Ast.~Letters, 13, 233 
\bibitem[\protect\citename{Gilfanov \& Sunyaev }1998]{gs98} Gilfanov M.,
         Sunyaev R., 1998, in preparation 
\bibitem[\protect\citename{Ghisellini, Haardt \& Matt }1994]{ghm94} 
         Ghisellini G., Haardt F., Matt G., 1994,
        \mnras, 267, 743 
\bibitem[\protect\citename{Illarionov et al.\ }1979]{ill79} Illarionov
        A., Kallman T.,  McCray R.,  Ross R., 1979, \apj, 228, 279  
\bibitem[\protect\citename{Inoue }1985]{ino85} Inoue H., 1985, \ssr, 40, 317
\bibitem[\protect\citename{Katz }1987]{katz87} Katz J.L,
        1987, \aap, 182, L19
\bibitem[\protect\citename{Koyama }1994]{koy94} Koyama K, 1994, New
        Horizon of X-ray Astronomy, FSS-12, 181, Univ. Acad. Press, Tokyo 
\bibitem[\protect\citename{Koyama et al.\ }1996]{koy96} Koyama K.,
        Maeda Y., Sonobe T., Takeshima T, Tanaka Y., Yamauchi S., 1996,
        Publ. Astron. Soc. Japan, 48, 249
\bibitem[\protect\citename{Lis \& Goldsmith }1989]{lg89} Lis D.C.,
        Goldsmith P.F., 1989, \apj, 337, 704
\bibitem[\protect\citename{Markevitch, Sunyaev \& Pavlinsky
        }1993]{msp93} Markevitch M, Sunyaev R., Pavlinsky, 1993, \nat,
        364, 40
\bibitem[\protect\citename{Matt, Perola \& Piro }1991]{mpp91} Matt G.,
        Perola G.C., Piro L., 1991, \aap, 247, 25
\bibitem[\protect\citename{Menten et al.\ }1997]{mre97} Menten K.M,
        Reid M.J., Eckart A., Genzel R., 
        1997, \apjl, 475, L111
\bibitem[\protect\citename{Milgrom }1987]{mil87} Milgrom M.,
        1987, \aap, 182, L21
\bibitem[\protect\citename{Morrison \& McCammon }1983]{mm83} Morrison
        R.,  McCammon D., 1983, \apj, 270, 119
\bibitem[\protect\citename{Nandra \& George }1994]{ng94} Nandra K., 
        George I.M., 1994, \mnras, 267, 974
\bibitem[\protect\citename{Rees }1988]{rees88} Rees M.J., 1988,
        \nat, 333, 523
\bibitem[\protect\citename{Pavlinsky, Grebenev \& Sunyaev
        }1994]{pav94} Pavlinsky M., Grebenev S., Sunyaev R., 1994,
        \apj, 425, 110 
\bibitem[\protect\citename{Predehl \& Tr\"{u}mper }1994]{pt94} Predehl
        P., Tr\"{u}mper J., 1994, \aap, 290, L29
\bibitem[\protect\citename{Rieke, Rieke \& Paul }1989]{rrp89} Rieke
        G.H., Rieke M.J., Paul A.E., 1989, 
        \apj, 336, 752 
\bibitem[\protect\citename{Stark et al.\ }1991]{sta91} Stark A.A.,
         Gerhard O.E., Binney J., Bally J., 1991
         \mnras, 248, 14p 
\bibitem[\protect\citename{Sunyaev et al.\ }1993]{smp93} Sunyaev R.,
        Markevitch M., Pavlinsky M., 1993, \apj, 407, 606 
\bibitem[\protect\citename{Sunyaev \& Churazov }1996]{sun96} Sunyaev
        R., Churazov E., 1996, Astronomy Letters, 22, 648
\bibitem[\protect\citename{Turner et al.\ }1997]{tur97} Turner M.J.L.,
        et al., 1997, 
        The Next Generation of X-ray Observatories: Workshop
        Proceedings, M.J.L. Turner \& M.G. Watson, eds., Leicester
        X-ray Astronomy Group Special Report, XRA97/02, p.165 
\bibitem[\protect\citename{Tanaka \& Shibazaki }1996]{tan96} Tanaka
        Y., Shibazaki N., 
        1996, Annual Review of Astronomy and Astrophysics, 34, 607.
\bibitem[\protect\citename{Vainshtein \& Sunyaev }1980]{vs80}
        Vainshtein L., Sunyaev R., 1980, Soviet~Ast.~Letters, 6, 673
\bibitem[\protect\citename{Vainshtein, Sunyaev \& Churazov }1998]{vsc98}
        Vainshtein L., Sunyaev R., Churazov E., 1998, Astronomy
        Letters, (in press).
\bibitem[\protect\citename{de Vicente et al.\ }1997]{vic97} de Vicente
        P., Martin--Pintado J., Wilson T.L., 1997, \aap, 320, 957
\bibitem[\protect\citename{Wise \& Sarazin }1992]{ws92} Wise M.W.,
        Sarazin C.L., 1992, \apj, 395, 387  
\bibitem[\protect\citename{White, Tananbaum \& Kahn }1997]{whi97}
        White N.E., Tananbaum H., Kahn S.M., 1997,
        The Next Generation of X-ray Observatories: Workshop
        Proceedings, M.J.L. Turner \& M.G. Watson, eds., Leicester
        X-ray Astronomy Group Special Report, XRA97/02, p.173 
\end{thebibliography}
\end{document}